\documentclass[twocolumn]{svjour3}          
\smartqed  
\usepackage[usenames]{color}
\usepackage[table]{xcolor}
\usepackage{soul}
\usepackage{textcomp}
\usepackage{multicol} 
\usepackage{multirow} 
\usepackage{xspace}
\usepackage{amsmath} 
\usepackage{graphicx}
\usepackage{breqn}
\usepackage{algorithm}
\usepackage{algorithmic}
\usepackage{amsfonts}

\usepackage{amssymb}

\newcommand{\data}[1]{{\textsl{#1}}}
\newcommand{\algo}[1]{{\textsf{#1}}}

\newcommand{\HL}[1]{\bfseries{#1}}

\def\T{\mbox{$\mathcal{T}$}} 
\def\G{\mbox{$\mathcal{G}$}} 
\def\V{\mbox{$\mathcal{V}$}} 
\def\E{\mbox{$\mathcal{E}$}} 

\def\L{\mbox{$\mathcal{L}$}} 

\journalname{SNAM}
\begin{document}

\title{Time-aware Analysis and Ranking of Lurkers in Social Networks
\thanks{An abridged version of this paper appeared in~\cite{ASONAM14}.}}


\author{Andrea Tagarelli \and Roberto Interdonato}


\institute{Dept. Computer Engineering, Modeling, Electronics,   and Systems Sciences. 
 University of Calabria, Italy\\ 
 \email{\{tagarelli,rinterdonato\}@dimes.unical.it}
}

\date{Received: date / Accepted: date}

\maketitle

\begin{abstract} 
Mining  the silent members of an online community, also  called lurkers, has been recognized as an   important problem that accompanies the extensive use of online social networks (OSNs). Existing solutions to the ranking of lurkers can aid understanding the lurking behaviors in an OSN. However,  they are limited to use only   structural properties of the static network graph, thus ignoring any relevant information concerning the time dimension.  Our goal in this work is to push forward research in lurker mining in a twofold manner:  (i) to provide an in-depth analysis of temporal aspects   that aims to unveil the behavior of lurkers and their relations with other  users, and (ii) to enhance existing methods for ranking lurkers by integrating different time-aware properties   concerning  information-production and  information-consumption actions.  Network analysis and ranking evaluation performed  on Fli\-ckr, FriendFeed and Instagram  networks 
allowed us to draw  interesting  remarks on both the understanding of lurking dynamics and on transient and cumulative scenarios of time-aware ranking.  
\end{abstract}

\keywords{lurking behavior \and time-aware lurker ranking \and LurkerRank algorithms \and newcomers \and inactive users \and responsiveness \and preferential attachment \and lurker trends and clustering \and topical evolution patterns}

\section{Introduction}
Lurking is a widely common behavior in online users, which is usually associated with definitions of nonparticipation, infrequent or occasional posting and, more generally, with observation, and bystander behavior~\cite{NonneckeP00,PreeceNA04}.  As a fundamental premise, it should be noted that lurkers should not be trivially regarded as totally inactive users, i.e., registered users who do not use their account to join an online community; rather, a lurker can be perceived as someone who gains benefit from other's information and services without significantly giving back to the online community. 

The main general reasons behind the multifaceted nature of this kind of user behavior  are well explained in social science, based on various motivational factors, such as environmental, commitment, quality requirements, and individual factors~\cite{Sun+14}.  
In general, lurkers represent an enormous potential in terms of social capital, because they acquire knowledge from the online community  but never or rarely let other people know their opinions. 
Lurking can indeed be expected or even encouraged because it allows users  to learn or improve their understanding of the etiquette of an online community before they can decide to provide a valuable contribution over time~\cite{Edelmann13}.  
Within this view, a major goal is to \textit{de-lurk} those users, i.e., to encourage lurkers to more actively participate in the online community life: indeed, even though a proper amount of lurkers is acceptable  for a large-scale social environment, too many individuals of that kind would impair the virality of the online community.  
 
 However,  a complete characterization of lurkers has represented a controversial issue in social  science and human-computer interaction research~\cite{Edelmann13}, which   has consequently posed several challenges in (quantitatively) analyzing lurking in online social networks (OSNs). 
   Despite the fact that lurkers represent  the large majority of members in an OSN, little research in computer science has been done that considers lurking as a valid  and worthy-of-investigation form of online behavior.  
In~\cite{SNAM14,ASONAM}, we fill a lack of knowledge on the opportunity of analyzing lurkers in OSNs, and on the important implications that the detection of lurkers can have on a deeper understanding of the feelings in an online  community.  
We   addressed the previously unexplored   problem of     ranking lurkers in an OSN, by introducing a topology-driven lurking definition and proposing a computational framework that offers various solutions to the ranking of lurkers. 
 However, a   limitation of the study in~\cite{SNAM14,ASONAM} is that it does not deal with temporal information to enhance the understanding and ranking of lurkers in an OSN.

Online social environments are highly dynamic systems, as individuals join, participate, attract, cooperate, and disappear across time. This clearly affects  the shape of the network both in terms of its social (followship) and interaction     graphs~\cite{jeong2003measuring,KumarNT06,leskovec2009meme,budak2011structural,lehmann2012dynamical,WilsonSPZ12,de2013analyzing,BerlingerioCGMP13}.   
Moreover, everybody agrees on the stance that users normally look for the most updated information, 
therefore the timeliness of users and their relations become  essential for evaluation~\cite{YuLL04,BerberichVW05,Smyth05,LangW13,JiangWWSHDZ13}. 
Like any other user,  lurkers as well may be  interested not only in the  authoritative sources of information, but also in the timely sources.

Research on temporal network analysis and mining strives to understand the driving forces behind the evolution of OSNs and what  dynamical patterns are produced by an interplay of various user-related dimensions in OSNs.
Dealing with the temporal dimension to mine lurkers appears to be even more challenging. Yet, it's also an emergent necessity, as users in an OSN  naturally evolve  playing different roles,   showing a stronger or weaker tendency toward lurking at different times.   
Moreover, as temporal dimension in an OSN is generally examined in terms of online frequency of the users, it's important to take into account that 
lurkers may  have unusual frequency of online presence as well as unusual frequency of interaction with other users.

\paragraph*{Contributions.} \ 
Our contributions in this work are twofold.    
First, we provide insights into the understanding of lurkers from various perspectives  along the time dimension in an OSN environment.  We conduct different stages of  temporal analysis of lurking behaviors, focusing on two macro aspects:      how lurkers relate to   other types of users in the network, and how patterns of lurking behaviors  evolve over time.  

Second,   we overcome the time-related limitation  of previous formulations of lurker ranking methods. 
To this purpose, we model   different temporal aspects concerning both the production and consumption of information, 
by introducing novel measures of  \textit{freshness} and \textit{activity trend},   at user  and at user-interaction level. 
These measures are   key ingredients in the proposed  time-aware lurker ranking methods, for which we  develop two approaches:  
a \textit{time-transient based ranking} approach, which is restricted to a particular snapshot graph of the network, and  a \textit{time-cumulative based ranking} approach, which  encompasses a      sequence of snapshots based on a time-evolving definition of freshness and activity functions.  

We structure  our work into seven research questions ({\bf Q1 - Q7}), which are summarized as follows. 
\begin{itemize}
\item Lurking is   often related to inactive behavior or to inexperienced usage of the network services at a given time. Therefore, we aim at unveiling whether and to what extent there exists  any correspondence between \textit{lurkers and zero-contributors} ({\bf Q1}), and between \textit{lurkers and newcomers} ({\bf Q2}). From a different perspective, we want also to understand whether lurkers create \textit{preferential relations with active users} ({\bf Q3}). 
\item  \textit{Responsiveness}, i.e., the willingness of a user to respond to other users, is a key criterion to measure behavioral dynamics of users in an OSN. We are hence interested in quantifying how frequently lurkers react to the postings of   other users ({\bf Q4}). 
\item  We investigate \textit{how lurking trends evolve over time} and how these can be characterized using a clustering framework ({\bf Q5}).   Moreover, by involving also the content dimension, we analyze the \textit{topical usage behaviors of lurkers} and their topic-sensitive evolution patterns ({\bf Q6}).  
\item We assess  the ability of our proposed \textit{time-aware lurker ranking algorithms} in providing improved solutions to the lurker ranking problem. We evaluate the impact of   the proposed   time-transient and time-cumulative based approaches  on the ranking performance, and also compare them with a state-of-the-art time-aware ranking algorithm~\cite{BerberichVW05} ({\bf Q7}).   
\end{itemize}

\paragraph*{Plan of the paper.} \  
  The remainder of this paper is organized as follows.  Section~\ref{sec:TLRsection}  provides first a short overview of our early work on    lurker detection and ranking in OSNs, then focuses on our proposal of time-aware LurkerRank methods. We answer to each of the above stated research questions in Section~\ref{sec:analysis}.  
  Section~\ref{sec:relatedwork}   discusses related work, and Section~\ref{sec:conclusion} concludes the paper.

\section{Time-aware Lurker Ranking}
\label{sec:TLRsection}

\subsection{LurkerRank at a glance}  
\label{sec:backgroundLR} 
In~\cite{SNAM14,ASONAM}, we developed  the first  formal computational  methodology for  lurker detection and ranking. We provided  well-principled definitions of lurking,  introduced a  network graph model oriented to the analysis and mining of lurkers, and defined methods to search and rank lurkers in an OSN.  

Our initial definition of lurking  relies solely on the topology information available in a OSN, modeled as a followship graph. Upon the assumption that   lurking behaviors build on the \textit{amount of information a user receives},  our key intuition is that  the strength of  a user's lurking status can be determined based on her/\-his  \textit{in/out-degree ratio} (i.e., followee-to-follower ratio), and of her/his neighborhood.  
 We report next the topol\-ogy-driven lurking definition from~\cite{SNAM14}:

\begin{definition}[Topology-driven lurking]\label{def:base}
Let   $\G = \langle \V, \E \rangle$ denote the directed graph representing  an OSN, with 
set of nodes (users) $\V$ and set of edges $\E$, whereby the semantics of any edge $(u, v)$ is that 
 $v$ is consuming information produced by $u$.  
A node $v$ with infinite in/out-degree ratio (i.e., a sink node) is trivially regarded as a lurker. 
A node $v$ with  in/out-degree ratio not below 1 shows a lurking status, 
whose strength is determined  based on:
\begin{description}
\item[\em Principle I:] \textit{Overconsumption}. The excess of informa\-tion-consumption over information-production.   The strength of $v$'s lurking status is   proportional  to its in/out-degree ratio. 
\item[\em Principle II:] \textit{Authoritativeness of the information received.}  The valuable amount of information received from  its in-neighbors. The strength of $v$'s lurking status is     proportional to the influential (non-lurking) status of the $v$'s in-neighbors.
\item[\em Principle III:] \textit{Non-authoritativeness of the information   produced.}  The non-valuable amount      of  information sent to   its out-neighbors. The strength of $v$'s lurking status is     proportional to the  lurking status of the $v$'s out-neighbors.
\end{description}  
\end{definition}

The above principles form the basis for three ranking functions that differently account for 
  the contributions of a node's in-neighborhood and out-neighborhood. We finally provided 
  a complete specification of our lurker ranking models in terms of PageRank-style methods.  
For the sake of brevity here, and throughout  this paper,   we will refer to only one of the  formulations described in~\cite{SNAM14,ASONAM}, 
which is that based on the full \textit{in-out-neighbors-driven lurker ranking}, hereinafter dubbed  simply as \textit{LurkerRank}. 

Given a node $u \in \V$, let us denote with  $B_u$ and $R_u$  the set of in-neighbors (i.e., backward nodes) and the set of  out-neighbors (i.e., reference nodes)  of $u$, respectively. The in-degree and  out-degree of $u$ are   denoted as $in(u) = |B_u|$ and $out(u)=|R_u|$, respectively. The following formula gives the LurkerRank $LR(v)$ for any node  $v$:
\begin{equation}\label{eq:LR}
LR(v) = d  [\mathcal{L}_{\mathrm{in}}(v)  \ (1+\mathcal{L}_{\mathrm{out}}(v))]  +   (1-d)/(|\V|)   
\end{equation}
where $\mathcal{L}_{\mathrm{in}}(v)$ is the in-neighbors-driven lurking function:
\begin{equation}
 \mathcal{L}_{\mathrm{in}}(v) =  \frac{1}{out(v)} \sum_{u \in B_v}   \frac{out(u)}{in(u)} LR(u) 
\end{equation}  
and $\mathcal{L}_{\mathrm{out}}(v)$ is the out-neighbors-driven lurking function:
\begin{equation}
\mathcal{L}_{\mathrm{out}}(v) =  \frac{in(v)}{\sum_{u \in R_v} in(u)} \sum_{u \in R_v}  \frac{in(u)}{out(u)} LR(u)  
\end{equation}  
Moreover,  $d$ is a damping factor ranging within [0,1], usually set to 0.85.  
To prevent zero or infinite ratios, the values of  $in(\cdot)$ and $out(\cdot)$ are Laplace add-one smoothed.


\subsection{Time-aware LurkerRank methods}
\label{sec:TLR}

In this section we describe our extensions to    LurkerRank  that account for   the temporal dimension when determining the lurking scores  of users in the network.   
We follow two   approaches based on different models of temporal graph: 

\textbullet \  \textit{Transient  ranking}, i.e.,  a measure of a user's  lurking score based on a time-static (snapshot) graph model;

\textbullet \ \textit{Cumulative ranking}, i.e., a measure of a user's  lurking score  that encompasses a given time interval (sequence of snapshots), based on a time-evolving graph model.   

The building blocks of our methods rely on the specification of the temporal aspects of interest, 
namely \textit{freshness} and \textit{activity trend}, both at user   and at user relation level. 
Freshness takes into account the timestamps of the latest information produced (i.e., posted) by a user, 
or   the timestamps of the latest information consumed by a user in relation to another user's  action.  Activity trend models how the users' posting actions   or the responsive actions vary over time.    
 These concepts will be elaborated on in the next section.

\newcommand{\Tau}{\mathcal{T}}
\subsubsection{Freshness and activity trend functions}

Users in the network   are assumed to perform actions and interact with each other over a timespan $\Tau \subseteq \mathbb{T}$.  The temporal domain $\mathbb{T}$ is conveniently assumed to be $\mathbb{N}$. Therefore, the time-varying graph of an OSN is seen as   a discrete   time system, i.e., the time is discretized at a fixed granularity (e.g., day, week, month).

\paragraph{Freshness.\ } 
Let $T \subseteq \Tau$ be a temporal subset of interest, being in interval notation of the form 
$T=[t_s, t_e]$, with $t_s \leq t_e$.  
For any time $t$, we define the \textit{freshness function} $\varphi_T(t)$ as:
\begin{equation}
 \varphi_T(t) =\begin{cases}
    1/\log_{2} (2+(t_e - t)), & \text{if $t \in T$}\\
    0, & \text{otherwise}.
  \end{cases}
\end{equation}
Function $\varphi_T(t)$ ranges within $[0,1]$. Note that we opt for a function with logarithmic decay to ensure, as $(t_e - t)$ gets larger, a slower decrease w.r.t. other decreasing functions with values in $(0,1]$---for instance, the graph of $\varphi_T(t)$ lies always above the graph of $2/(1+\exp(t_e-t))$, or  of $1/(1+(t_e-t))$.

Given a  user  $u$, let  $T_u$ be the set of time units at which $u$ performed  actions in the network. The \textit{freshness} of $u$ at a given temporal subset of interest $T$  is defined as:  
\begin{equation}
f_T(u) = \max \{ \varphi_{T}(t), \ t \in T_u \ \mbox{s.t.} \ t_s \leq t \leq t_e \}  
\end{equation}
Note that $f_T(u)$ is always   defined and   positive, for all $t \in T$.  Higher values of $f_T(u)$ correspond to more recent activities of $u$ w.r.t. $T$.

\paragraph{Activity trend.\ } 
The second aspect we would like to understand is the activity trend of a user. 
Let us first denote with
\begin{equation}\label{eq:activity-series}
S_u = [(x_1, t_1), \ldots, (x_n, t_n)]
\end{equation}
the time series representing the \textit{activity} of user $u$ over  $T_u$.  
For every pair $(x, t) \in S_u$, $x$ denotes the number of $u$'s actions at time $t$.

In order to model the temporal evolution of the activity of a user $u$, we employ the \textit{Derivative time series Segment Approximation} (DSA)~\cite{GulloPTG09}  and apply it to the user's activity time series  $S_u$.  DSA  is able to represent a time series into a concise form which is designed to capture the significant variations in the time series profile. 
For any given time series $S_u$ of length $n$, DSA produces a new series $\tau$ of $h$ values, with $h \ll n$.   
The main steps performed by DSA are summarized as follows:  

\begin{itemize}
\item
\textit{Step 1 - Derivative estimation}: $S_u$ is transformed into $S'_u$, where each value $x \in S_u$ is replaced by its first
derivative estimate.
\item 
\textit{Step 2 - Segmentation}:  the derivative time series $S'_u$ is partitioned into $h$  
variable-length segments. Each of the segments aggregates subsequent data
values having very close derivatives, i.e., it represents a subsequence of values with a specific trend.
\item
\textit{Step 3 - Segment approximation}: each of the segments in $S'_u$ is mapped to an angular value $\alpha$, which collapses   information on the average slope  within the segment.
 \end{itemize}

The DSA series $\tau_u$ is of the form $\tau_u = [(\alpha_1, t_1), \ldots$ $\ldots, (\alpha_h, t_h)$, such that 
$\alpha_j = \arctan(\mu(s_j))$ and $t_j = t_{j-1} + l_j$, with $j=[1..h]$, where $s_j$ is the $j$-th segment, $l_j$ its  length, and $\mu(s_j)$ the mean of its points. 

 As a post-processing step, the values $\alpha_j$  of the DSA sequence  $\tau_u$ are normalized within [0,1] by deriving the values    $\hat{\alpha_j} =  \alpha_j/\pi + 1/2$. In this way,   an increasing (resp. decreasing)  trend of activity will correspond to  a value within  $(0.5,1]$ (resp. $[0,0.5)$).  Therefore, we define the \textit{activity trend} of user $u$ (over the whole interval $T_u$) as the time sequence:
\begin{equation}
a(u) = [(\hat{\alpha}_1, t_1), \ldots, (\hat{\alpha}_h, t_h)]
\end{equation}
Given a temporal interval of interest $T \subseteq T_u$, the activity trend of $u$ w.r.t. $T$ corresponds to the subsequence $a_T(u)$ of $a(u)$ that   fits $T$.  It is also useful to define the \textit{average activity} of $u$ over  $T$, denoted by $\overline{a_T}(u)$, as the average of the $\hat{\alpha}$ values within $a_T(u)$.

\paragraph{Freshness and activity trend of interaction.\ } 
The notions of freshness and activity trend for individual users are here extended to model the interaction of any   two users $u, v$ at a given time $t$, which corresponds to the directed  edge $(u, v)$  (also here denoted as $u \rightarrow v$) in the snapshot graph  containing  $t$. 
The rationale here is that the more recent is an interaction (or the more increasing is its activity trend), the stronger should be the relation $u \rightarrow v$.  Recall that $(u,v)$ means that $v$ is consuming information  at time $t$  produced by $u$. 

Let us denote with $P_u = \{p_1, \ldots, p_k\}$   the set of information-production actions of $u$, and with  $T_u(P_u)=\{ t_{p_1}, \ldots, t_{p_k}\}$ the  associated timings.  Moreover, 
let $C_{u \rightarrow v}$ be a set of triplets  $(t_{p_i}, t_{c_j}, x_{c_j})$, such that   $t_{p_i} \in T_u(P_u)$, 
and $t_{c_j}, x_{c_j}$ denote the time and the frequency, respectively, at which $v$ consumed  the $u$'s post $p_i$. 
Note that we have used subscripts $_p$ and $_c$ to mean ``production'' and ``consumption'', respectively. 

According to the above  formalism, we define the \textit{freshness of interaction} $u \rightarrow v$ w.r.t. $T$ as the maximum freshness over the sequence of pairs \textit{(production-time, consumption-time)}    in $T$:  
\begin{multline}
f_T(u, v) = \max \{  \varphi_{[t_p, t_c]}(t_p), \\
    s.t. \ \ \exists (t_p, t_c,\_) \in C_{u \rightarrow v} \wedge t_s \leq t_p, t_c \leq t_e    \} 
\end{multline}

Analogously, we  define the activity trend of interaction, based on the DSA model previously used to define the activity trend of user. To this end, given the interaction $u \rightarrow v$, we consider  the time series $S_{u,v}$ representing  pairs $(x,t)$, where $x$ denotes the number of actions at time $t$ performed by $v$ in response to a specific post by $u$. 
Then, we compute the \textit{activity trend of interaction} $u \rightarrow v$, denoted with $a_T(u, v)$, as the result of the application of  DSA to the time series $S_{u,v}$.  The definition of $\overline{a_T}(u,v)$ is analogous to $\overline{a_T}(u)$.

\subsubsection{The time-static LurkerRank algorithm}

Our first formulation of time-aware LurkerRank is based on a  time-static graph model, which contains one single snapshot of the network. Our key idea   is to capitalize on the previously proposed functions of freshness and activity  to define  a time-aware weighting scheme  that determines both the strength of the productivity of a user and the strength of the interaction between any two  users  linked at a given time.  To this purpose, we introduce two real-valued, non-negative coefficients $\omega_f, \omega_a$ to control the importance of  the freshness and  the activity trend in the weighting scheme.   

Given a temporal interval of interest $T$, and coefficients $\omega_f, \omega_a$, we define the   function $w_T(\cdot)$  
in terms of the user freshness and average activity   calculated for any  user $v \in \V$: 
\begin{equation}\label{eq:node-weight}
w_T(v) = \begin{cases} 
\frac{\omega_f f_T(v) + \omega_a \overline{a_T}(v)}{\omega_f + \omega_a}, & \mbox{if } f_T(v) \neq 0, \overline{a_T}(v) \neq 0 \\
 f_T(v), & \mbox{if } f_T(v) \neq 0, \overline{a_T}(v) = 0 \\ 
 1, & \mbox{otherwise}\end{cases}
\end{equation}
By default, the two coefficients are set uniformly as   $\omega_f = \omega_a = 0.5$. 
If  $T_u$ is contained into $T$ (i.e., $f_T(v) \neq 0$) and the average activity is zero,   the  $w_T$ value will coincide to the freshness value, which is strictly positive;  otherwise, if  $f_T(v) = 0$, the  $w_T$ value  will equal one.  
It should be noted that $w_T$ will hence be 1 if either the freshness and  average activity are maximum or $T$ is not relevant to the timespan over which the user has been active: this is admissible in our theory since we want to exploit information about the user activity only if this is available in a given time interval. 
The rationale behind the $w_T$ value assigned to a vertex $v$ is to add a multiplicative factor that is inversely (resp. directly) proportional, otherwise neutral,  to the size of the in-neighborhood
$in(v)$ (resp. size of the out-neighborhood $out(v)$) in the formulation of  our time-static LurkerRank algorithm. 

Analogously to  $w_T(\cdot)$, we  define     the   function $w_T(\cdot, \cdot)$  in terms of the    freshness and average activity   of interaction calculated for    any $u,v \in \V$  such that $(u,v) \in \E$, as follows: 
\begin{equation}\label{eq:edge-weight}
w_T(u,v) = \begin{cases} 
\frac{\omega_f f_T(u,v) + \omega_a \overline{a_T}(u,v)}{\omega_f + \omega_a}, & \! \mbox{if } f_T(u,v) \neq 0, \\ & \overline{a_T}(u,v) \neq 0 \\
 f_T(u,v), & \mbox{if } f_T(u,v) \neq 0, \\ & \overline{a_T}(u,v) = 0 \\ 
 0, & \mbox{otherwise}\end{cases}
\end{equation}
 Compared to Eq.~(\ref{eq:node-weight}), note that the expression in  Eq.~(\ref{eq:edge-weight})   holds zero  if the freshness is zero.  This will be clear as we will show the use of $w_T(\cdot,\cdot)$ in an exponentially negative smoothing term that is present in the definition of our time-static LurkerRank algorithm.

We are now ready to provide our formulation of the \textit{time-static LurkerRank} algorithm, hereinafter denoted as  \algo{Ts-LR}, which involves both functions $w_T(\cdot)$ and $w_T(\cdot, \cdot)$ above defined. Time-static LurkerRank shares with the basic LurkerRank formulation the way the  in-neighbors-driven lurking term is combined with the  out-neighbors-driven lurking term, that is, for any user $v \in \V$ and temporal interval of interest $T$:  
\begin{equation}\label{eq:tsLR}
Ts\mbox{-}LR_T(v) = d  [\mathcal{L}_{\mathrm{in}}(v)  \ (1+\mathcal{L}_{\mathrm{out}}(v))]  +   (1-d)/(|\V|)   
\end{equation}
However,  the in-neighbors-driven lurking function $\mathcal{L}_{\mathrm{in}}(v)$ is now defined as:\footnote{Note that, for the sake of simplicity, we have omitted the subscript $T$ in the freshness and activity trend functions, in the weighting function as well as in the $in$ and $out$ functions, since the reference   interval of interest $T$ is   assumed clear from the context. Analogously, we override  the function symbols $\mathcal{L}_{\mathrm{in}}(v)$ and $\mathcal{L}_{\mathrm{out}}(v)$ given in Eq.~(\ref{eq:LR}), since  they will be never  referenced out of the \algo{Ts-LR} setting.}
\begin{multline}
 \mathcal{L}_{\mathrm{in}}(v) =  \frac{1}{w(v) out(v)}     \exp\left(-\sum_{u \in B_v} w(u,v)\right)  \\     \sum_{u \in B_v}   \frac{out(u)}{in(u)} Ts\mbox{-}LR_T(u) 
\end{multline}  
and  the out-neighbors-driven lurking function  $\mathcal{L}_{\mathrm{out}}(v)$ as:
\begin{multline}
\mathcal{L}_{\mathrm{out}}(v) =  \frac{in(v)}{w(v) \sum_{u \in R_v} in(u)}   \exp\left(-\sum_{u \in R_v} w(v,u)\right)   \\  \sum_{u \in R_v}  \frac{in(u)}{out(u)} Ts\mbox{-}LR_T(u)  
\end{multline}

\subsubsection{The time-evolving LurkerRank algorithm}
The time-static LurkerRank can work only on a subset of relational data that are  restricted to a particular subinterval of the network timespan. Therefore,  information on the sequence of events concerning users' (re)actions is lost as relations are aggregated into a single snapshot.  
To overcome this issue, we define here an alternative  formulation of time-aware LurkerRank that is able to model, for each   user $v$,  the potential accumulated   over a time-window of the contribution  that  each in-neighbor $u$ had to the computation of the lurking score of $v$. 

\paragraph*{Cumulative freshness and activity functions.} \ 
We begin with the definition of  a \textit{cumulative scoring function}  which forms the basis for each of the   subsequent  functions that will apply to the previously defined freshness and average activity at user and interaction level.  
Intuitively, this cumulative scoring function ($\mathrm{g}_{\leq}$)  should be defined at any time $t \in T$ to aggregate all values of a function $\mathrm{g}$ (defined in $T$) computed at times $t' \in \T$ less than or equal to $t$, following an exponential-decay model: 
\begin{equation}
\mathrm{g}_{\leq}(t) \propto \mathrm{g}(t) + \sum_{t' < t}  (1-2^{t'-t}) \mathrm{g}(t') 
\end{equation}

Let the timespan $\mathcal{T}$ of the network graph be partitioned in consecutive sub-intervals  $T_1, T_2, \ldots, T_i, \ldots = [t_0, t_1], (t_1, t_2], \ldots, (t_{i-1}, t_{i}] \ldots$.  
 The generic cumulative scoring function $\mathrm{g}_{\leq}(\cdot)$ has a straightforward translation in terms of user-freshness: 
 if $t_i$ corresponds to the end-time of the span of interest whose latest sub-interval is $T_i$,  
 we define  the \textit{cumulative user-freshness} function applied to any user $u$ to  integrate (with exponential decay) all user-freshness values individually obtained at each sub-interval preceding $t_i$:
\begin{equation}\label{eq:cfu}
cf_{T_i}(u)= f_{T_i}(u) + \sum_{t_k < t_i}  (1-2^{t_k-t_i}) f_{T_k}(u) 
\end{equation}
Our  \textit{cumulative  user-activity} function, we denote with $ca_{T_i}(\dot)$, has similar form. Formally, for every $u \in \V$, we have:
\begin{equation}\label{eq:cau}
ca_{T_i}(u)=\overline{a_{T_i}}(u) + \sum_{t_k < t_i}  (1-2^{t_k-t_i}) \overline{a_{T_k}}(u) 
\end{equation}

The definition of cumulative freshness of interaction, $cf_{T_i}(u,v)$,  and cumulative  activity  of interaction, $ca_{T_i}(u,v)$,  at each $T_i$, and for every $(u, v) \in \E$, follow  intuitions analogous to Eq.~(\ref{eq:cfu}) and Eq.~(\ref{eq:cau}), respectively.  

The values yielded by the above defined four functions of cumulative freshness and activity, at user as well as at interaction level,  are then normalized and multiplied by the corresponding  information in the transient model, that is, for every $u \in \V$:    
\begin{equation}\label{eq:cf-norm}
cf_{T_i}^{~~\prime}(u) =  \frac{ cf_{T_i}(u) } { \max_j  cf_{T_j}(u)}   f_{T_i}(u)
\end{equation}
\begin{equation}\label{eq:ca-norm}
ca_{T_i}^{~~\prime}(u)=  \frac{ ca_{T_i}(u) } { \max_j  ca_{T_j}(u)}   \overline{a_{T_i}}(u)
\end{equation}
The  user-interaction function counterparts have analogous form to Eq.~(\ref{eq:cf-norm}) and Eq.~(\ref{eq:ca-norm}).   
Our motivation for adopting a (multiplicative) combination of a normalized cumulative freshness/activity function with a transient freshness/activity function, is that we want to ensure that the freshness/activity information cumulated through times preceding a target time $T_i$ will be valued w.r.t. the actual contribution (in terms of freshness/activity) that the user provides in the OSN at given time $T_i$.
  
The \textit{time-evolving LurkerRank} algorithm, hereinafter denoted as \algo{Te-LR}, follows a formula that is analogous to the \algo{Ts-LR}. However, \algo{Te-LR} adopts new  weighting functions, we denote as $cw_T(\cdot)$ and $cw_T(\cdot, \cdot)$, which have the following properties:
\begin{itemize}
\item  they have analytical form that is    identical to $w_T(\cdot)$, given in Eq.~(\ref{eq:node-weight}),  and $w_T(\cdot, \cdot)$, given in Eq.~(\ref{eq:edge-weight}), respectively;
\item they are defined, at user level, in terms of the   functions $cf_{T_i}^{~~\prime}(\cdot)$ and $ca_{T_i}^{~~\prime}(\cdot)$ (given in Eq.~(\ref{eq:cf-norm}) and Eq.~(\ref{eq:ca-norm})), and at user-interaction level, in terms of the   functions $cf_{T_i}^{~~\prime}(\cdot, \cdot)$ and $ca_{T_i}^{~~\prime}(\cdot, \cdot)$.
\end{itemize}

\section{Temporal Analysis of Lurkers}
\label{sec:analysis}
 
We present here our multi-faceted, temporal analysis of lurkers  along the time dimension. In Section~\ref{sec:outline}, we   frame   seven  research questions, which span different problems of interest to our study.   We describe the data that will be used for our evaluation in Section~\ref{sec:data-description}.  Then, Sections~\ref{sec:Q1}--\ref{sec:Q7} will contain our answer to each of the stated questions.

\subsection{Outline of research questions}
\label{sec:outline}

Our study of lurkers and lurking behaviors across time is built upon seven research questions, which are stated as follows.

 {\bf Q1}: \textit{Do lurkers match  zero-contributors?} 
Definitions of lurking are often related to nonposting behavior. 
Our first research question is aimed at gaining insights into the correspondence between inactive users and lurkers over time. Inactive users are here intended as  ``zero-contributors'', i.e.,   users who have never posted or provided a comment/favorite-mark.
 
  {\bf Q2}: \textit{Do lurkers match   newcomers?} 
Lurking   can depend on a temporary status of learning the etiquette of the community and the proper usage of the  services provided by an OSN. This also relates  to newcomers, i.e., users that have started to participate in some activity.  
Therefore, analogously to our first research question, we also analyzed the relation  between newcomers   and lurkers over time. 

 {\bf Q3}: \textit{Do lurkers create preferential relations with active users?}
In the third research question, our goal is to unveil the dynamics of the binding between lurkers and active users, and how this relates to the popularity of the active users.

 {\bf Q4}: \textit{How frequently do lurkers respond to the others' actions?} 
Lurkers can  show a limited amount of activity  in response to others' contributions to the community life. We are   interested in measuring the distribution of time latency that occurs to observe repeated  actions by a user in response to his/her followees  (i.e., comments, or favorite-marks).
 
 {\bf Q5}: \textit{How do  lurking trends evolve?} 
Our fifth research question focuses on how lurking trends change over time, how they can be grouped together, and wheth\-er characteristic  patterns may arise to indicate different profiles of lurkers.

 {\bf Q6}: \textit{How do topical interests of  lurkers evolve?} 
We are also interested in exploring topic-sensitive evolution patterns of lurking behavior. We analyze the topical usage of lurkers, how lurkers change their topical patterns, and whether these changes might differ from   those of the other users. 
 
{\bf Q7}: \textit{Can time-aware models improve the ranking of lurkers?}  
 In our final research question, we investigate the impact of using our proposed  time-transient and    time-cumulative ranking models on the quality of lurker ranking solutions.  
 We   provide a quantitative analysis of results obtained by   our developed time-aware LurkerRank algorithms, with respect to a data-driven evaluation of the ranking performance. We also offer a comparison  with a state-of-the-art time-aware ranking algorithm~\cite{BerberichVW05}.

\subsection{Data}
\label{sec:data-description}

We used data from Flickr, FriendFeed and Instagram networks  to conduct our analysis.  
A major motivation  underlying our  data selection is that  we wanted to use datasets that have been previously studied in research and that  contain timestamped   information on the    activities of  users and their relationships, including \textit{followships}, \textit{comments}, or \textit{like/favorite-markings}.  
 \data{Flickr} dataset  was originally collected in 2006-2007 and used in~\cite{mislove08wosn,cha2009www}, \data{FriendFeed}   refers to the latest (2010) version of the dataset studied in~\cite{CelliLMPR10}, while \data{Instagram}   is our   dump recently crawled in 2014, whose user interaction  network   was used in~\cite{ASONAM14}.   \data{Flickr} and \data{FriendFeed} have  also been selected to be consistent with our previous analysis  of lurkers~\cite{SNAM14}.  
We refer the reader to the original works that used \data{Flickr} and \data{FriendFeed}, and to our submission support page available at  \textit{http://uweb.dimes.} \textit{unical.it/tagarelli/timelr/}        
 for    the description of the \data{Instagram} dump.

 \begin{table}[t!]
\caption{Main structural characteristics of the evaluation network datasets.}
\centering
\scalebox{0.75}{
\begin{tabular}{|l|c|c|c|c|c|c|}
\hline
\emph{data} & \textit{\# nodes} & \textit{\# links} &  \textit{avg}  &  \textit{avg} & \textit{clust.} & \textit{assorta-}  \\
  &   &  &  \textit{in-deg.}  &  \!\!\textit{path len.}\!\! & \textit{coef.} & \textit{-tivity}  \\
\hline
\hline
 \multicolumn{7}{|c|}{\textit{averages over time-varying snapshot graphs}}   \\
\hline
\data{Flickr-social}  &   1,889,102	 &   25,265,343 & 13.25 & 4.41  & 0.108 & 0.009\\
\hline
\data{Flickr} & 215,429  &  	1,483,462 &   6.85 &	4.69  & 0.025 & -0.013 \\
\hline
\data{FriendFeed}  &  6,962  & 	64,509  &  5.15 & 	5.89 &  0.071  & -0.043 \\
\hline
\data{Instagram}  &   10,353  &  31,215 & 2.94	 &   5.83  & 0.083 &  0.217 \\
\hline 
\hline
 \multicolumn{7}{|c|}{\textit{full (static) social graphs}}   \\
\hline
\data{Flickr}   &  2,302,925   & 33,140,018  &  14.39 & 4.36 & 0.107 & 0.015  \\
 \hline
\data{FriendFeed}     & 493,019    &  19,153,367   &   38.85  &  3.82  &  0.029  &  -0.128   \\
 \hline 
\data{Instagram}     & 54,018    &  963,833   &   17.85  &  4.50  &  0.048  &  -0.067   \\
\hline  
\end{tabular}
}
\label{tab:data}
\end{table}

 Note that our selected   datasets are rather heterogeneous in terms of features concerning user relationships: \data{Flickr} contains    timestamps of   34.7M   favorite markings assigned to the uploaded photos, and also contains (inferred) timings on the   user subscriptions. In \data{Instagram}, every link  between $v$ (follower) and $u$ (followee)  is annotated with the number and timestamp of the $v$'s comments to  media posted by $u$ (about 2M comments and 1.7M likes).  Analogous to \data{Instagram} is the situation in  \data{FriendFeed} but for information concerning likes ($\approx$230K) and comments ($>$687K) to posts.  

Table~\ref{tab:data} summarizes main structural characteristics of the network datasets we used in our evaluation. The table is organized in two subtables. The upper subtable contains  statistics on timestamped \textit{snapshot graphs} averaged over the network-specific timespan, which was binned at  month  level; more precisely, in order to have uniformly-sized snapshots, we aggregated them on a 28-days (i.e., 4 weeks) basis.      
Note also that all snapshots refer to \textit{interaction} subgraphs except the first row in the table which corresponds to the timestamped followship (social) subgraphs of \data{Flickr}.    
The timespans covered by the datasets are 7 months for \data{Flickr} (2006/11/02 -- 2007/05/17 for the followship subgraphs and  2006/09/08 -- 2007/03/22  for the interaction subgraphs), 7 months for  \data{FriendFeed} (2010/04/09 -- 2010/09/30), and 20 months for \data{Instagram} (2012/06/28 -- 2013/12/18). 
The lower subtable  contains statistics on the full (i.e., static) social graphs of \data{Flickr}, \data{FriendFeed} and \data{Instagram}.

\begin{figure}[t!]
\centering
\includegraphics[width=0.48\textwidth]{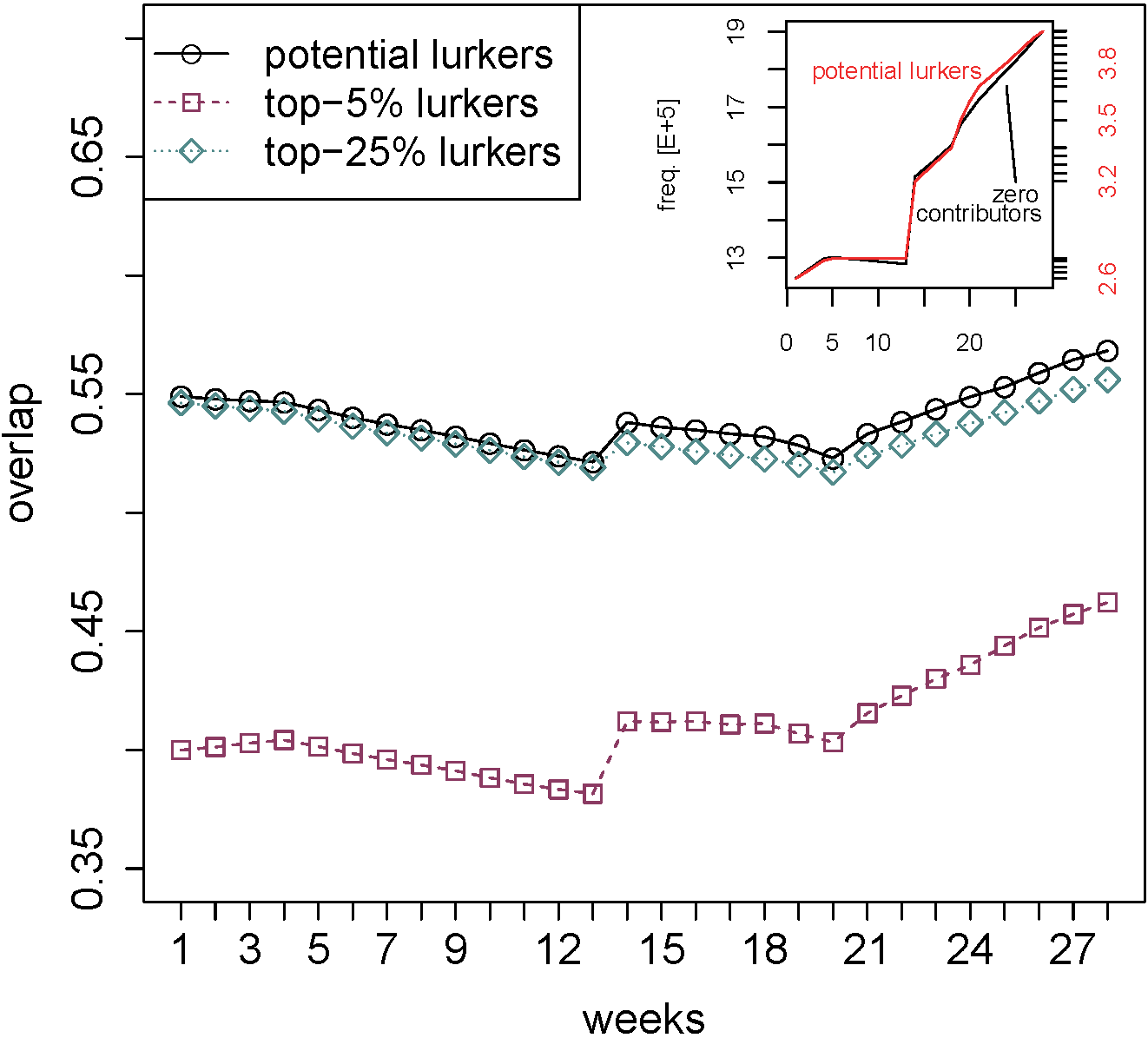}
\caption{Overlap ratio of zero-contributors against potential lurkers and top-ranked lurkers: distributions over weekly snapshots of the \data{Flickr} network. The inset shows the weekly distributions of zero-contributors and potential lurkers.}
\label{fig:overlap-zc}
\end{figure}

 \begin{figure*}[t!]
\centering
\begin{tabular}{ccc}
\includegraphics[width=0.3\textwidth]{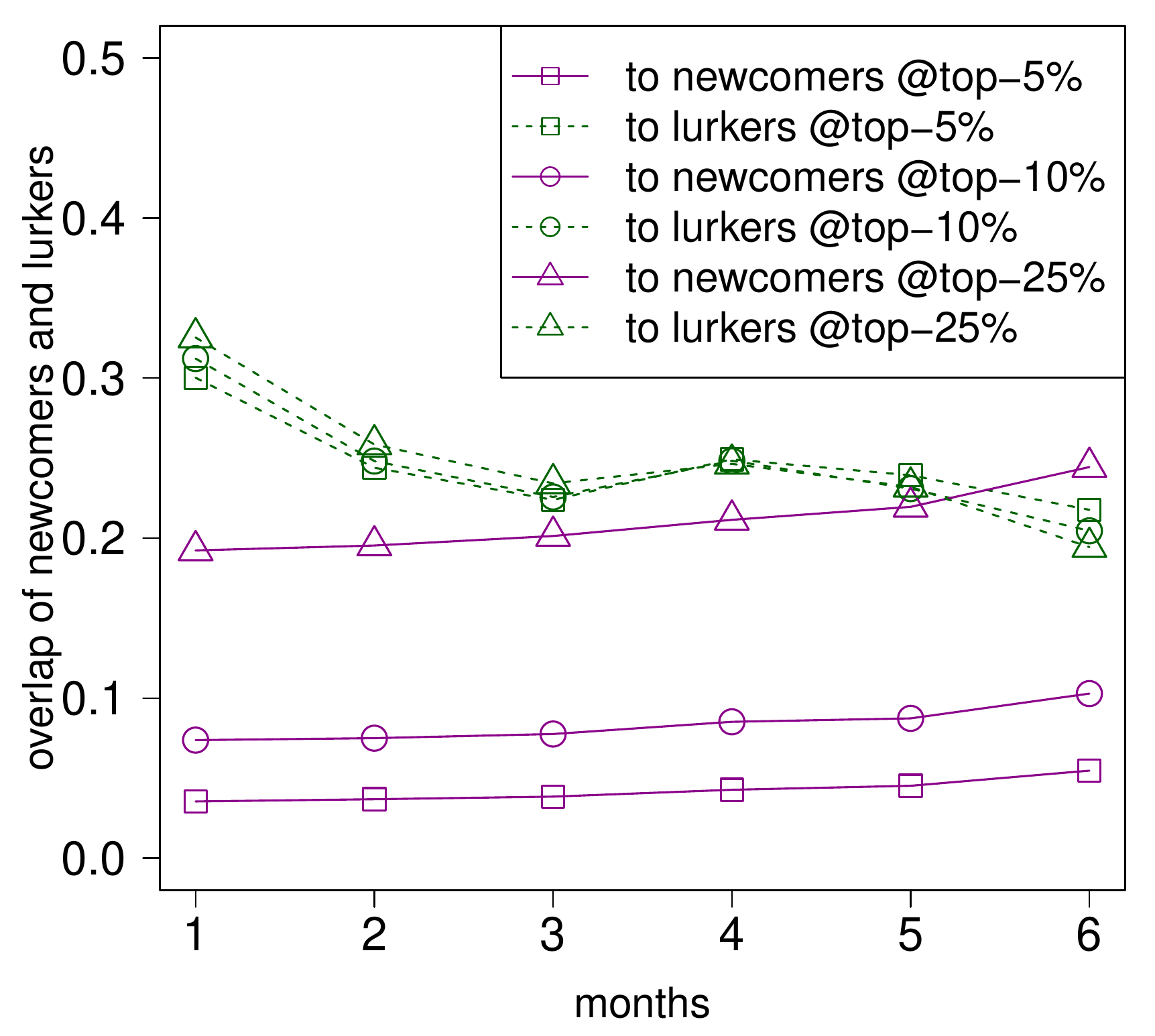} &  
\includegraphics[width=0.3\textwidth]{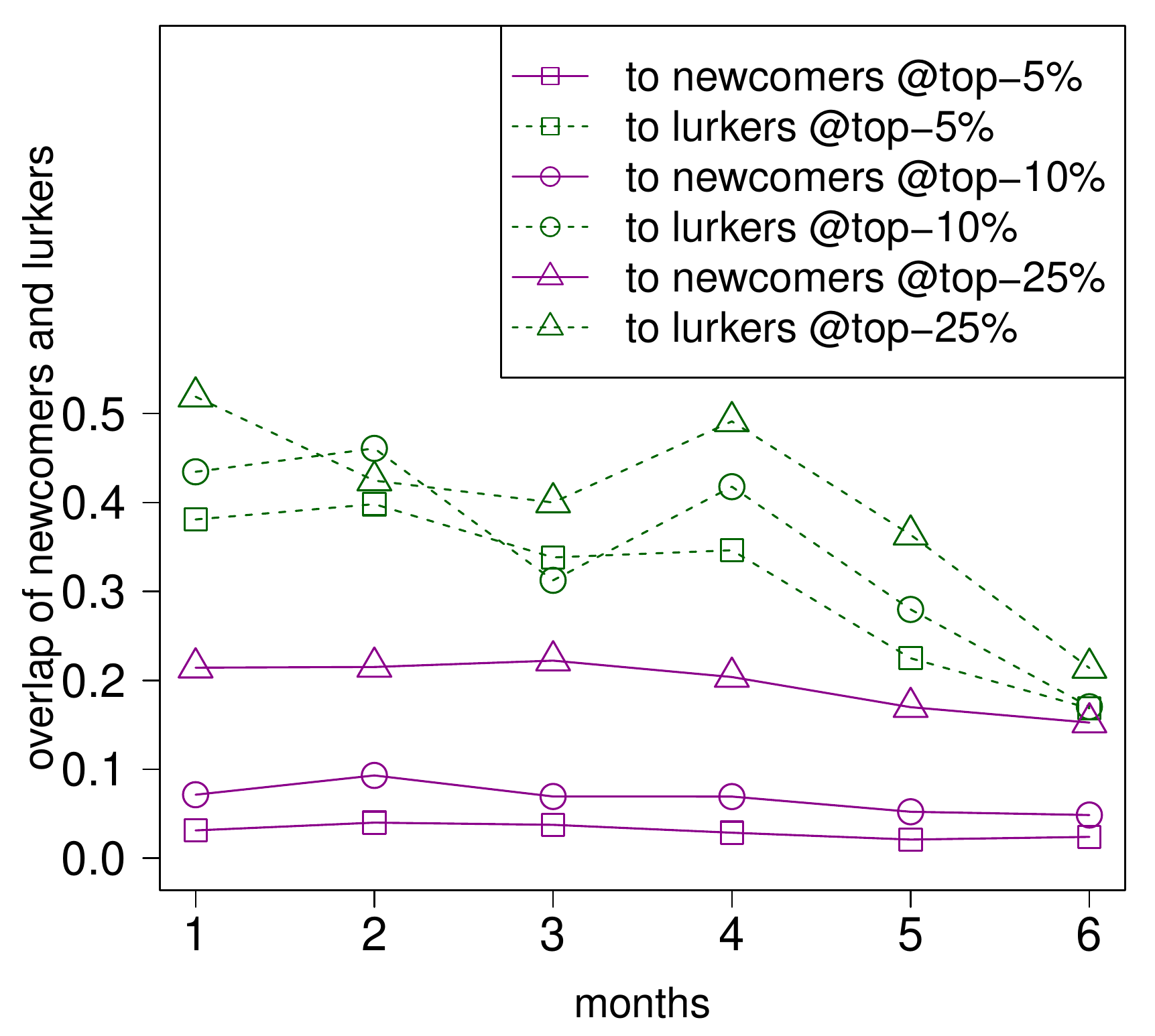} &  
\includegraphics[width=0.3\textwidth]{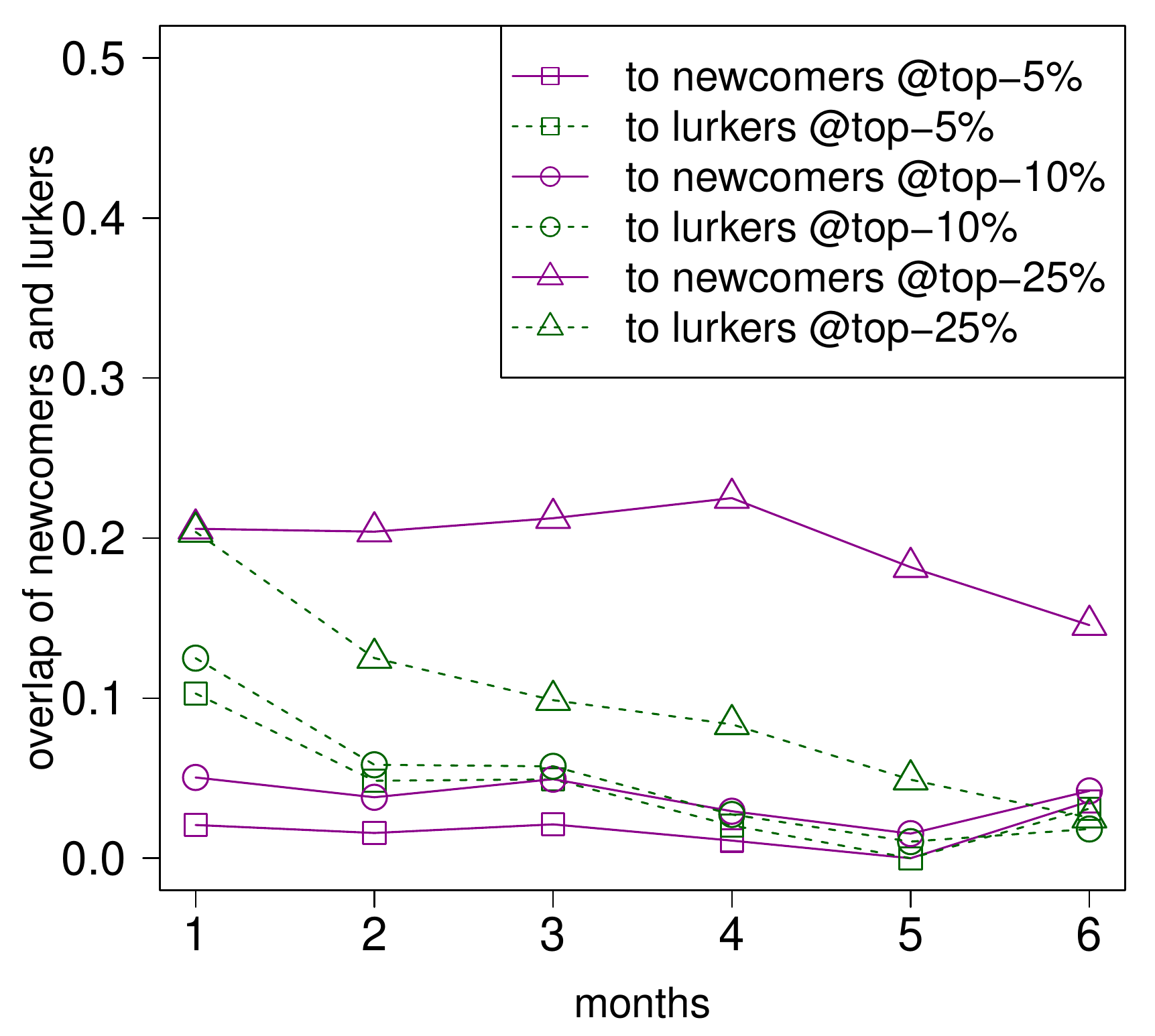}\\
(a) \data{Flickr} & (b) \data{FriendFeed} & (c) \data{Instagram}\\
\end{tabular}
\caption{Overlap ratio of newcomers and top-ranked lurkers in monthly snapshots.}
\label{fig:newcomers}
\end{figure*}

\subsection{Lurkers vs. inactive users}
\label{sec:Q1}

Our first question (\textbf{Q1}) focuses on the relation between lurkers and inactive users, also referred to as zero-con\-tributors. 

To answer this question, we initially analyzed how much the set of zero-contributors overlaps with the set of users having an in/out-degree ratio higher than one, here dubbed ``potential lurkers''. When considering the static picture of a network dataset,  
one remark is that the set overlap between zero-contributors and potential lurkers may vary from 12\% (favorite-based interaction network in \data{Flickr}) to 72\% and 95\% (comment-based interaction networks in \data{FriendFeed} and \data{Instagram}, respectively). Moreover, since the relative difference in size of the two sets can  vary from one dataset to another, we also computed the overlap ratio w.r.t. the set of potential lurkers, which was found to be   57\% on \data{Flickr}, 62\% on \data{Instagram}, and 96\% on \data{FriendFeed}. There are hence  clues that the overlap (or overlap ratio) would be relatively smaller when favorite/like interactions are taken into account, that is,  
potential lurkers are more likely to behave similarly to inactive users when activity is regarded in terms of commenting.

We further investigated how the relation between inactive and lurking users evolves over time. In this analysis, we also included  the set of top-ranked users obtained    by  our LurkerRank.    Figure~\ref{fig:overlap-zc} shows the temporal trends of overlap ratios w.r.t. potential lurkers, top-5\% and top-25\% ranked lurkers, on \data{Flickr}. 
Interestingly, the overlap ratios remain rather unaffected over time, despite   the jump in frequency at the 14-th week (displayed in the inset).  The distribution of top-5\% ranked lurkers is always above the other two series  (up to 0.15), which in turn roughly match. 
 Note   that in the inset, the distributions of potential lurkers and zero-contributors actually   follow close trends, although they are scaled differently (on one order of magnitude).

\begin{figure*}[t!]
\centering
\begin{tabular}{cccc}
\hspace{-3mm}
\includegraphics[width=0.24\textwidth]{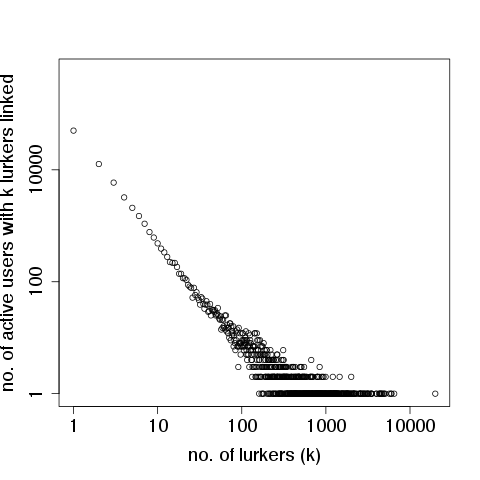} & \hspace{-3mm}
\includegraphics[width=0.24\textwidth]{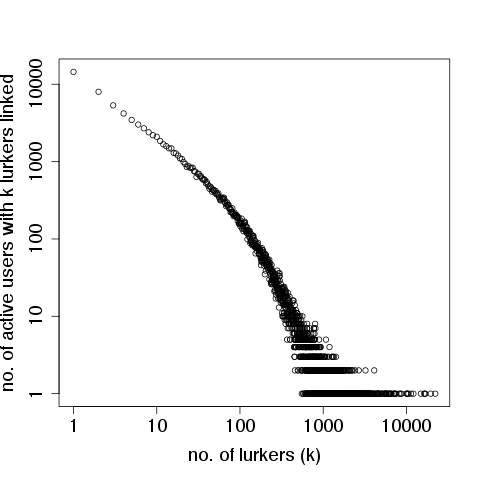} &
\hspace{-3mm}
\includegraphics[width=0.24\textwidth]{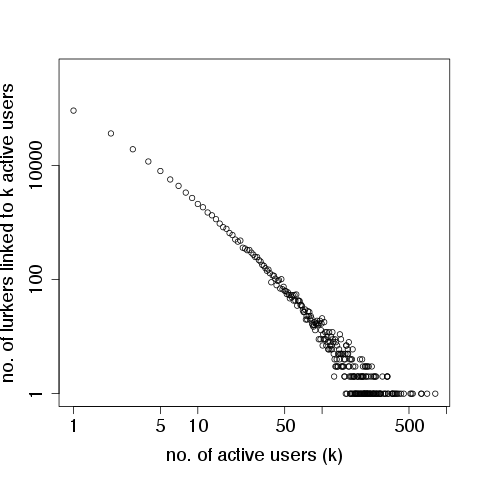} & 
 \hspace{-3mm}
\includegraphics[width=0.24\textwidth]{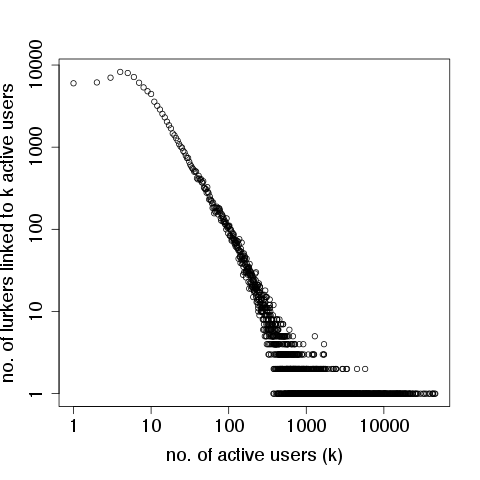}\\
(a) \data{Flickr} &  (b) \data{FriendFeed} & 
(c) \data{Flickr} &  (d) \data{FriendFeed} 
\end{tabular}
\caption{Distribution of  active users as a function of the lurkers-followers, (a) and (b), and distribution of  lurkers as a function of the active users-followees, (c) and (d).}
\label{fig:powerlaw-static}
\end{figure*}

\subsection{Lurkers vs. newcomers}
\label{sec:Q2}

Similarly to the previous analysis, in our second research question  (\textbf{Q2}), we investigated  whether and to what extent lurkers and    newcomers can overlap   at any given time  in our evaluation networks. 

 To this end, we assumed that  a user is regarded as a newcomer  at time $t$ if, at any time $t' < t$, s/he was not involved in any interaction with other users, while lurkers were identified at each time $t$.  Figure~\ref{fig:newcomers} shows, for each dataset and relating top-LurkerRank   solutions at 5\%, 10\% and 25\%, two series over a six-month timespan: the fraction of newcomers that were recognized as lurkers (solid lines) and the fraction of lurkers that were also newcomers (dashed lines).  
 
In the case of  interactions as \data{Flickr} favorite-markings actions, shown in  Fig.~\ref{fig:newcomers}(a),  
we observe that the fraction of lurkers matching newcomers varies  from about 30\% down to 20\%, 
following the same trend over the timespan regardless of the top-\% selected from the LurkerRank  solution; 
by contrast, the trend of the fraction of newcomers matching lurkers is more  constant (and slightly increasing)  over the timespan, achieving values below 10\%, for top-5\% and top-10\% lurkers, and around 20\% for top-25\% lurkers.  

Considering a comment-based interaction scenario, shown in  Fig.~\ref{fig:newcomers}(b)-(c), again  the trend of the fraction of newcomers matching lurkers looks roughly constant over time.   
As concerns  the fraction of lurkers matching newcomers, it is within 50-20\% in \data{FriendFeed}, but below 10\% on average in \data{Instagram}. 

 We tend to believe that the difference in matching (between lurkers and newcomers) among the various scenarios might be explained due to inherent characteristics of an OSN, rather than to the type of interaction (as previously observed in the evaluation of inactive users versus lurkers).  Nevertheless, we would like to point out that our research objective in comparing lurkers with newcomers is consistent with previous research focused on the analysis of newcomers' behavior  in an OSN: in fact, as found by Burke et al.~\cite{BurkeML09}, newcomers' behavior can be explained by examining how they   tend to be engaged in content production activities by \textit{observing} their friends' actions. This is nothing less than a    form of Bandura's \textit{observational learning}~\cite{Bandura}, i.e.,  learning through being given access to the learning experiences of other users; as widely studied in social science and human-computer interaction, observational learning and lurking are related to each other~\cite{Edelmann13,Sun+14}.

\subsection{Preferential attachment}
\label{sec:Q4}

Our third research question ({\bf Q3}) focuses on understanding  whether    relations  between lurkers and the active users they are linked to can be explained in terms of power law and preferential attachment.  
To this purpose, we selected the set of lurkers and the set of active users respectively from the top and the bottom of the LurkerRank   solution.     

We first investigated whether the probability of observing active users with a certain degree of attached lurkers, and vice versa, can be predicted by a power-law.  
Figure~\ref{fig:powerlaw-static} shows the  distribution of lurkers  as a function of the degree of attached active users, and also  for  the distribution of active users, obtained on the \data{Flickr} and \data{FriendFeed} followship graphs, using the  top-25\% and bottom-25\% of the LurkerRank   solution.   
We computed the best fit of a power-law distribution to the observed data, and assessed  the statistical significance of the fitting by a Kolmogorov-Smirnov test. 
From the figure it can be noted that the plots follow a power-law behavior. The exponents of the fitted power-law distributions are        
1.725 ($x_{min}=1$) and 1.363 ($x_{min}=1$)  for \data{Flickr} (Fig.~\ref{fig:powerlaw-static}(a) and (c), resp.), 
2.015 ($x_{min}=315$) and 1.679 ($x_{min}=99$) for \data{FriendFeed} (Fig.~\ref{fig:powerlaw-static}(b) and (d), resp.).   
In all cases, the power-law fitting is statistically significant, with  Kolmogorov-Smirnov test statistic (resp. $p$-value) of 0.0236 (resp. 0.8006) for Fig.~\ref{fig:powerlaw-static}(a), 0.0396 (resp. 0.7662) for Fig.~\ref{fig:powerlaw-static}(c), 0.0516 (resp. 0.9946) for Fig.~\ref{fig:powerlaw-static}(b), and 0.0546 (resp. 0.9161) for Fig.~\ref{fig:powerlaw-static}(d).

Our main goal to answer question \textbf{Q3} is  to try explaining the relation between lurkers and active users in terms of preferential attachment, that is, we    hypothesize that  lurking connections are attached preferentially to active users that already have a large number of connected lurkers.      
Following the lead of~\cite{mislove08wosn}, we studied two separate cases of  ``attachment'', which differently rely on a user's in-neighborhood or out-neighborhood. However, in our context, such a type of analysis becomes more complicated since nodes (and their neighborhood) must be selected according to their different status as either lurker or active user. Intuitively,  two cases of preferential attachment can be considered, namely:   new connections received   by active users for any $k$ lurkers, and   new connections produced  by lurkers for any $k$ active users.

\begin{figure}[t!]
\centering
\begin{tabular}{cc}
\hspace{-3mm}
\includegraphics[width=0.24\textwidth]{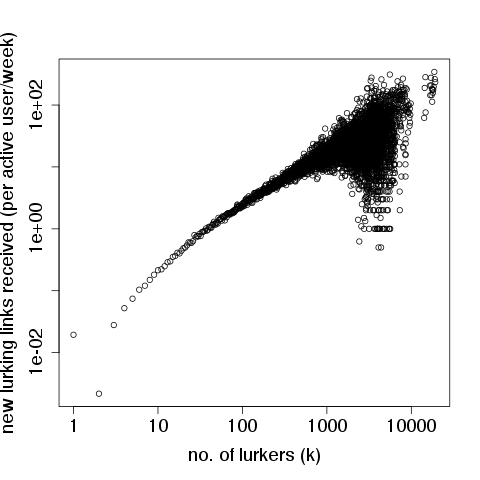} & \hspace{-3mm}
\includegraphics[width=0.24\textwidth]{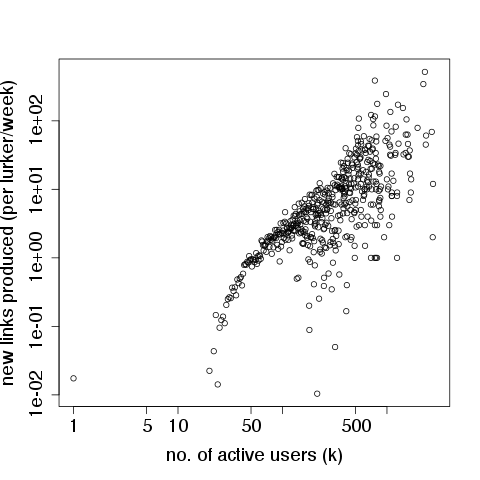}  
\end{tabular}
\caption{Timestamped followship-based evaluation of preferential attachment between lurkers vs. active users. New connections are detected for each weekly-aggregated network, on  \data{Flickr}.}
\label{fig:pref-attach}
\end{figure}

We initially investigated the two cases of preferential attachment  according to the timestamped followship information in a network.   
Figure~\ref{fig:pref-attach} shows results obtained on \data{Flickr}, averaged per user and per week, for each $k$.    
It can be noted that the number of lurkers shows a good linear correlation with the average number of   new links received by  active users  (left-hand side of Fig.~\ref{fig:pref-attach}):  
the least-squared-error linear fit has a slope of 0.00836, which means that on average active users  receive  per week one new connection from lurkers for every 120 lurker-followers   that they already have. 
By contrast,  given a  correlation of -0.11, no   linear trend exists   when studying the new connections produced by lurkers for any $k$ active users (right-hand side of Fig.~\ref{fig:pref-attach}).  
Therefore,   it is unlikely that  lurkers  following  a high  number of active users  will   create new connections towards other active users.

\begin{figure}[t!]
\centering
\begin{tabular}{cc}
\hspace{-3mm}
\includegraphics[width=0.24\textwidth]{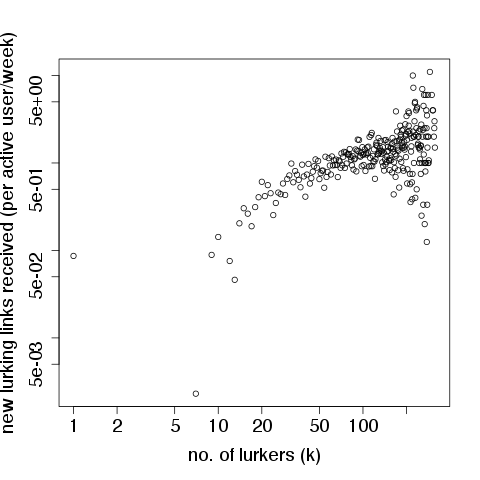} & \hspace{-3mm}
\includegraphics[width=0.24\textwidth]{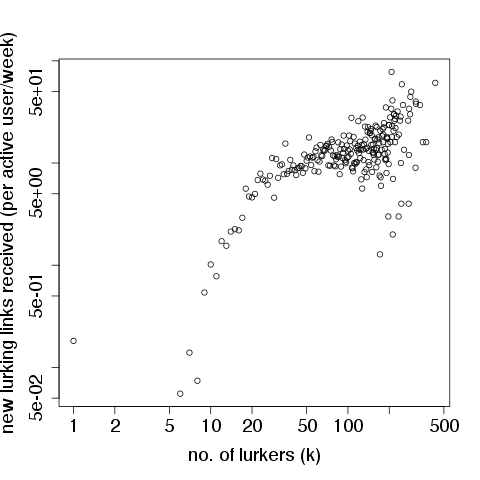}  \\
(a) \data{Instagram} & (b) \data{FriendFeed}
\end{tabular}
\caption{Timestamped interaction-based evaluation of preferential attachment between lurkers vs. active users. New connections are detected for each weekly-aggregated network.}
\label{fig:pref-attach1}
\end{figure}

We further explored the preferential attachment evaluation between lurkers and active users focusing on timestamped interaction information in a network.  
Specifically, we considered  user interaction based on likes or comments in \data{Instagram},  and on  comments in \data{FriendFeed}.  
 Figure~\ref{fig:pref-attach1} shows results obtained on the two datasets that  concern the correlation between the number of lurkers ($k$) and the average number of new links received by active users for any given $k$, on a weekly basis.  
Correlation is moderate (0.34 on \data{Instagram},  0.52 on \data{FriendFeed}), while, in terms of  least-squared-error linear fit, the two distributions have  a slope of 0.00570 (\data{Instagram}) and 0.06585 (\data{FriendFeed}), which correspond to having one new interaction (i.e., posted comment)  from lurkers per active user and   week   for every 176 and 15, respectively,  lurkers that have already interacted.  
However,  compared to the analogous situation on weekly-aggregated \data{Flickr} followship networks  in the left-hand side of Fig.~\ref{fig:pref-attach}, both the distributions in Fig.~\ref{fig:pref-attach1} have lower correlation and also   lower  size, which can be explained since   temporal information about user interactions (i.e., likes/comments) in both networks is relatively sparse with respect to that about followships.  
In effect, by aggregating interaction relationships on a monthly basis, correlation increases (0.47 on \data{Instagram},  0.66 on \data{FriendFeed}), along with a decrease of the amount of preferential attachment (one new interaction  from lurkers per active user and  month for every 51 and 4 interacting lurkers on \data{Instagram} and   \data{FriendFeed}, respectively).    
Finally, concerning new connections produced  by lurkers for any $k$ active users, we observed very sparse distributions with null correlation, on both datasets and     regardless of the temporal grain of the aggregated networks.  This means that lurkers, which  have a higher number of active users as recipients of their likes/comments, are not more prone to  have new interactions with  other active users.

 \begin{figure}[t!]
\centering
\includegraphics[width=0.3\textwidth]{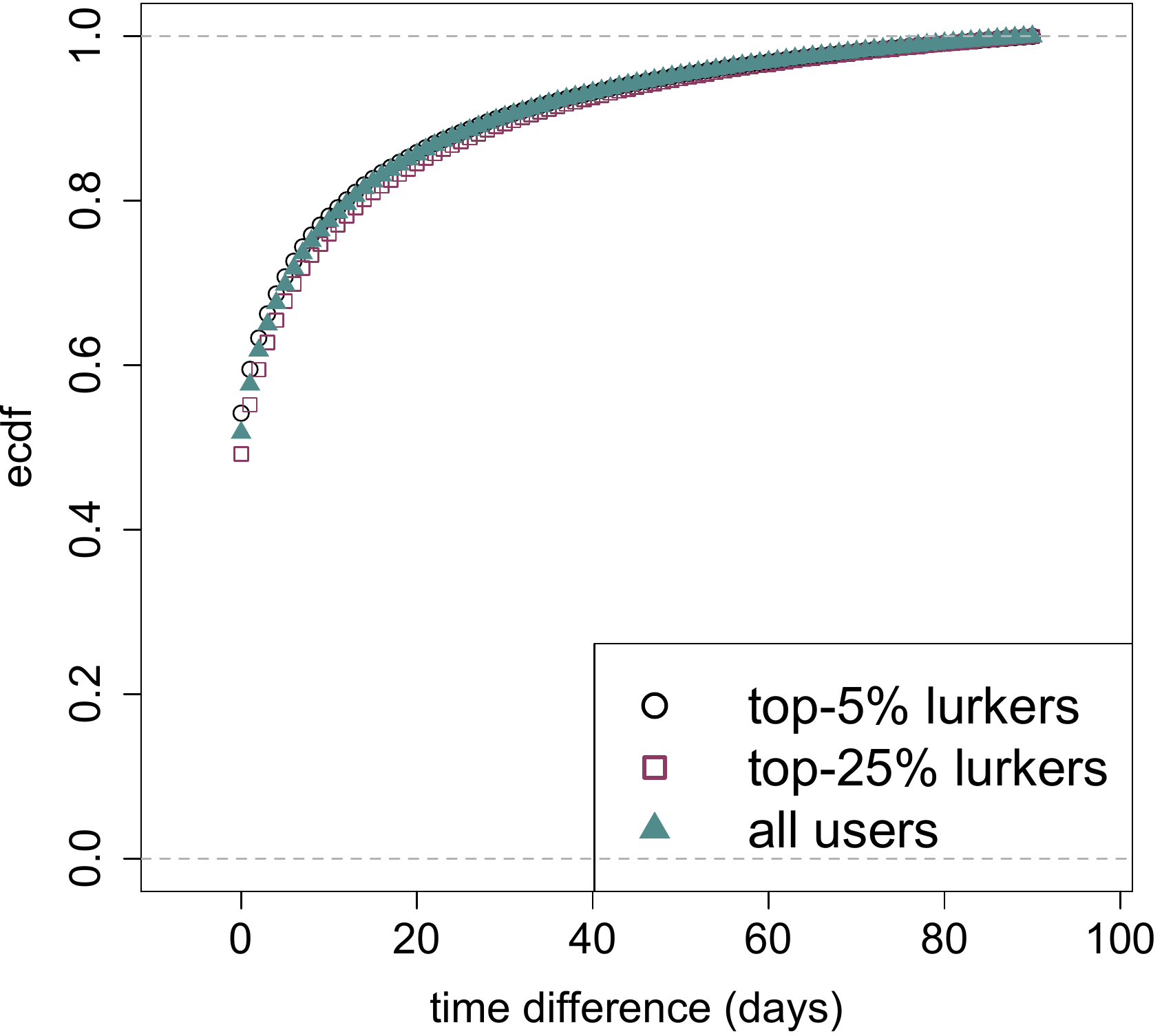} \\
(a) \data{Flickr} \\ \vspace{2mm}
\includegraphics[width=0.3\textwidth]{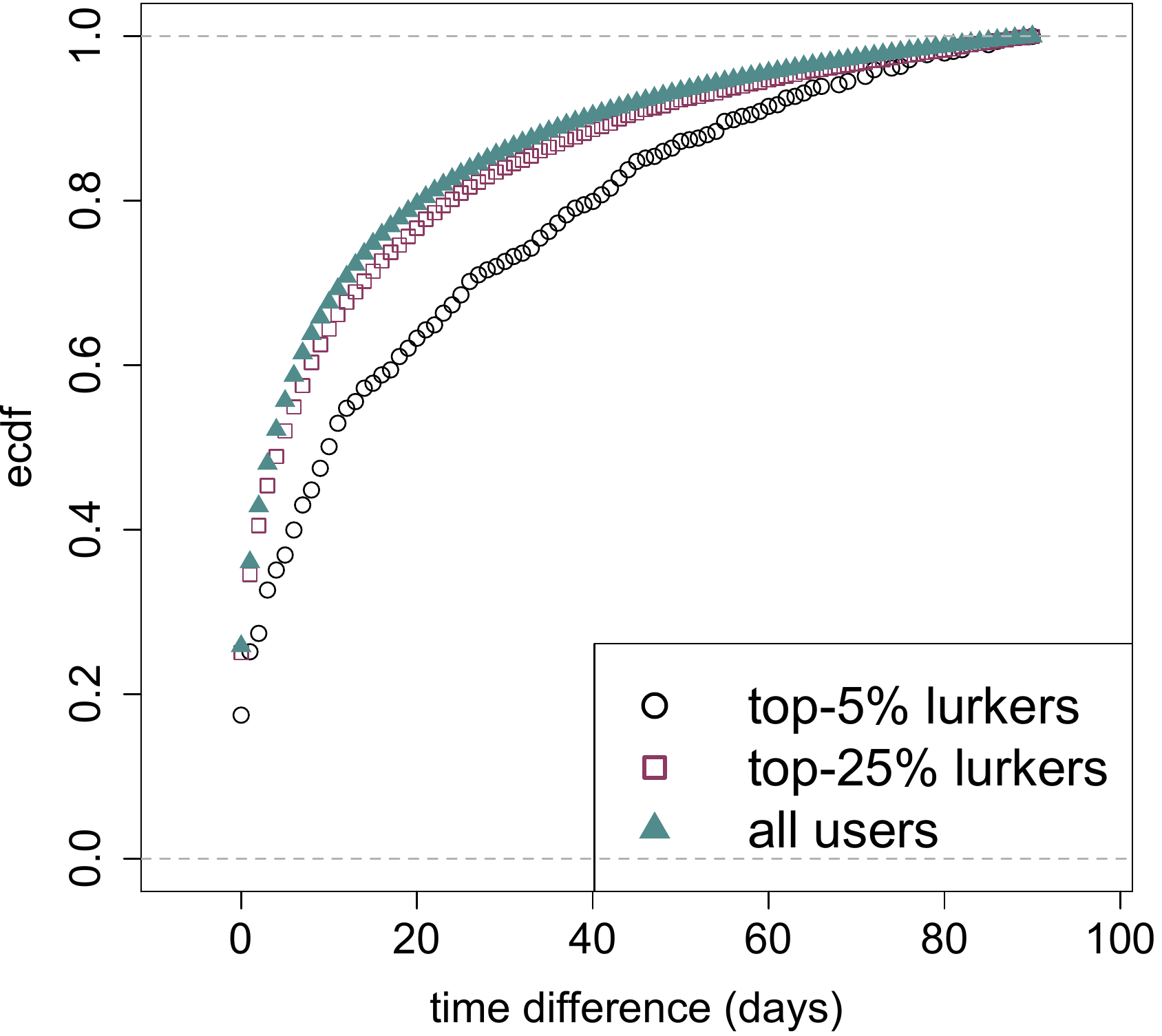} \\ 
(b) \data{Instagram}\\
\caption{Responsiveness frequency:  empirical cumulative distribution function (ecdf) plots of user reaction latency (in days), based on  favorites in \data{Flickr}   and  comments in \data{Instagram}. (Best viewed in color.)} 
\label{fig:timediff}
\end{figure}

\begin{figure}[t!]
\centering
\includegraphics[scale=0.48]{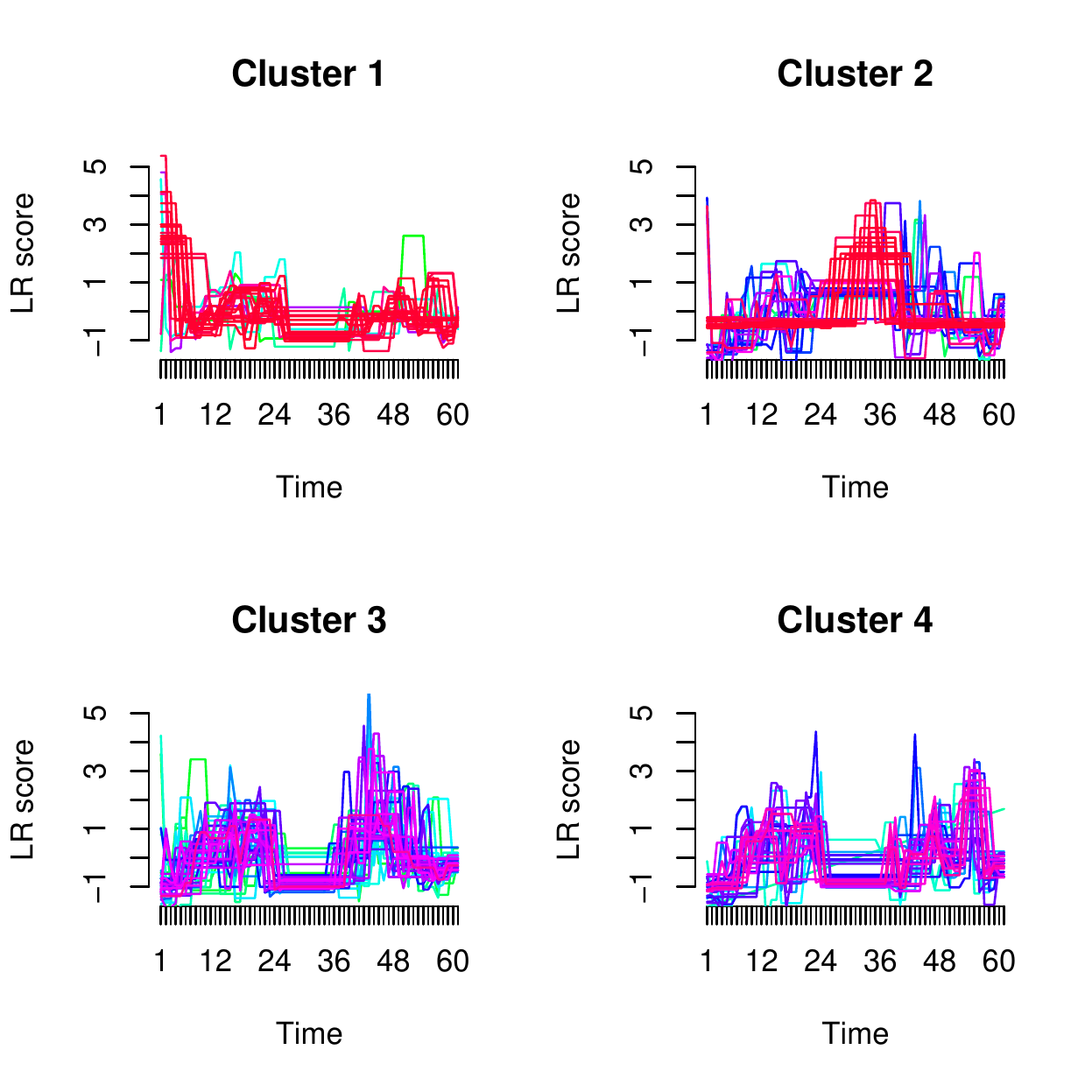}  \\
(a) \data{FriendFeed} \\
\includegraphics[scale=0.48]{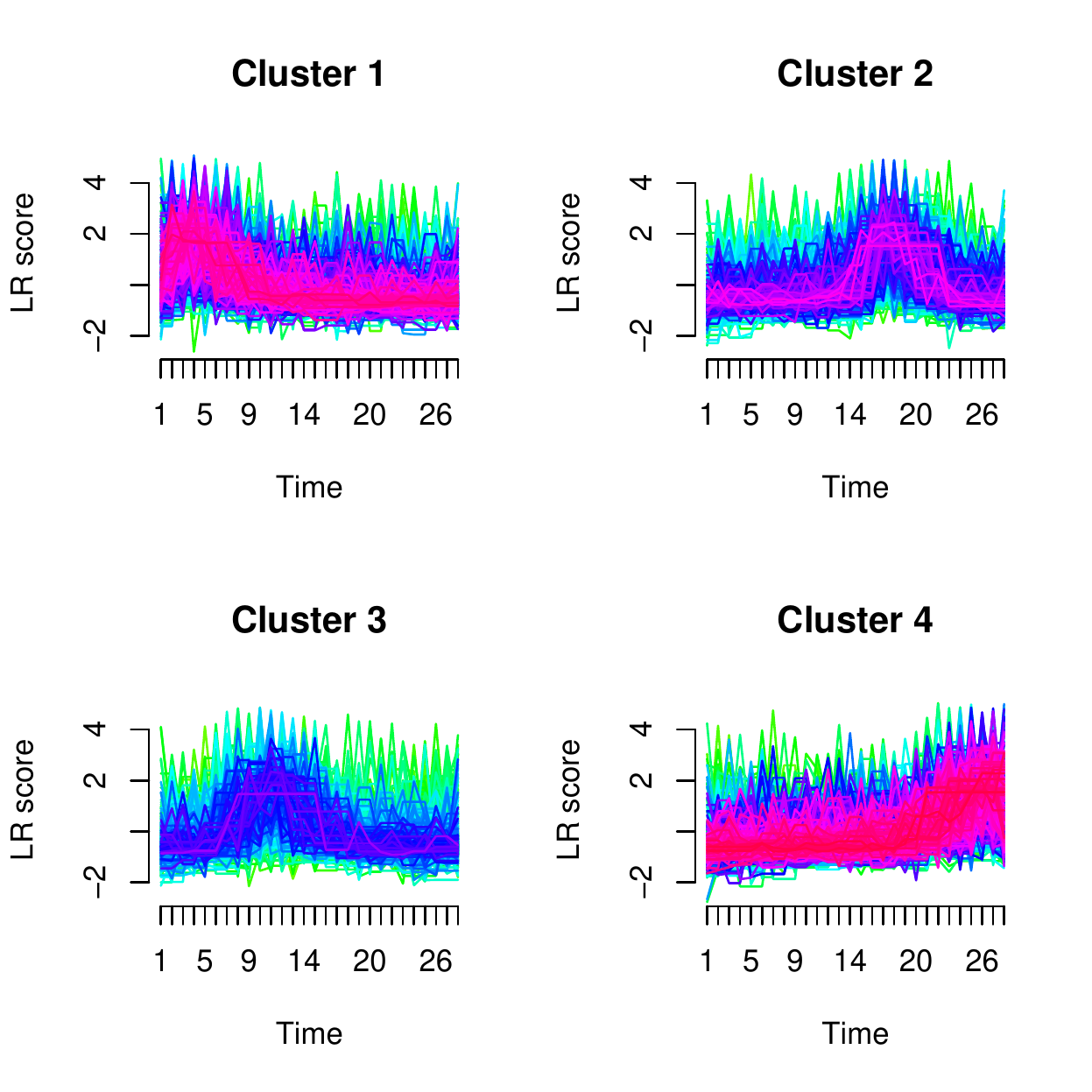}  \\
(b) \data{Flickr} \\
\includegraphics[scale=0.48]{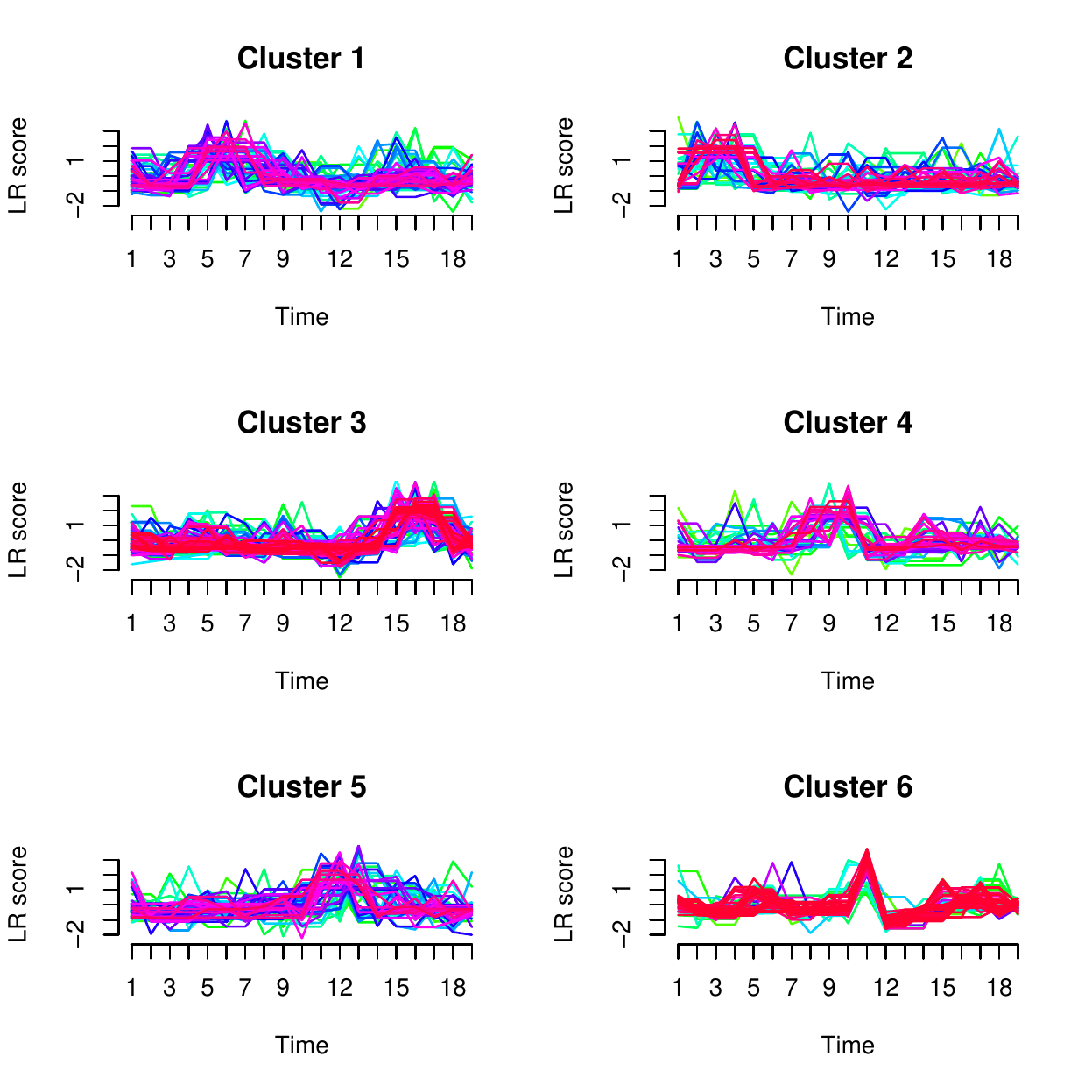}  \\
 (c) \data{Instagram}
\caption{Clustering of the time series representing LurkerRank scores in time-evolving graph networks: 
(a) \data{FriendFeed}   daily snapshots built on like+comment relations, 
(b) \data{Flickr}   weekly snapshots built on favorite relations, 
and (c) \data{Instagram}   monthly snapshots built on comment relations.  
Warmer colors correspond to series with higher cluster-membership.  (Best viewed in color.)}
\label{fig:ts-clustering}
\end{figure}

\subsection{Responsiveness}
\label{sec:Q3}

Concerning question  \textbf{Q4}, 
we aim at estimating   how frequently lurkers react to the postings of   other users. 

We examined the distribution of time  differences (in days) 
 between any two consecutive  responsive actions  made by a user w.r.t. a post created by  her/his followees.    
Figure~\ref{fig:timediff} shows  the  empirical cumulative distribution functions  over the first 90 days, for comments on  \data{Instagram} and for favorites on \data{Flickr}. Each of the  plots in the figure compares the distributions obtained for top-5\% and top-25\% lurkers with the distribution corresponding to all users in the network.  

We observe that  the lurkers' responsiveness generally takes   several days, or weeks, although the latency between   any two consecutive responsive actions may significantly vary in the two networks.   
 Focusing on the 80\% of responses (i.e., 0.8 on the $y$-axis of the plots), the latency is up to 18 days in \data{Flickr},  with no evident difference regarding the fraction of top-ranked lurkers considered; by contrast,  in \data{Instagram},   the top-25\% lurkers have an average responsiveness of more than three weeks, which takes even longer (40 days) in the case of   top-5\% lurkers.   
 Moreover, compared   with the responsiveness of   all users, lurkers tend to react more slowly, up to 20 days more in \data{Instagram}; however, the gap with respect to all users is only of few days in both networks when the fraction of top-ranked lurkers is large (25\%).  
Thus,  more time-consuming responsive actions, like comments in \data{Instagram}, would explain not only the increase in the  the lurkers'  responsiveness but also  the relative difference with  the generic case of all users.

\begin{table*}[t!]
\caption{Summary of LDA-learned topics in our \data{Instagram} dataset.}
\label{tab:topics}
\resizebox{\textwidth}{!}{
\begin{tabular}{|c|c|c|c|}
\hline
\multirow{2}{*}{LDA topic ids} &  topic-set  & \multirow{2}{*}{main  descriptors (i.e., media tags) of topic-set} & subnetwork-\\
                      &   label   &  &   induced size \\
\hline\hline
\multirow{2}{*}{0, 6, 10}  & \multirow{2}{*}{\textsf{nature}} & sky, sunset, whpflowerpower, whpsignsoftheseason, clouds, nature, landscape, & \multirow{2}{*}{8,185} \\
 & & sea, beach, flowers, water, trees, hinking, summer, fall, autumn   & \\
\hline
\multirow{2}{*}{12, 14} & \multirow{2}{*}{\textsf{architecture}} &  whpstraightfacades,  architecture, building, instaworld_shots,  streetphotography, & \multirow{2}{*}{2,884} \\ 
 & & spain, madrid, paris, france, london, sicily, design,  arquitectura,  youmustsee  &  \\
\hline
13 & \textsf{fun} &  love,  me,  swag,  lol,  fun,  like,  awesome,  cool,  happy, food   & 1,314\\
\hline
16 & \textsf{pets} &  whppetportraits, cats, caturday, catstagram, dog, cute, pets, kitty, catsofinstagram, petsofinstagram   & 3,124 \\
\hline
\multirow{2}{*}{19} & \multirow{2}{*}{\textsf{video}} & whpmovingphotos, whpreplacemyface, whpbigreveal, whpfilmedfromabove, instavideo,  & \multirow{2}{*}{3,062} \\
 &   &   video, whpmovingportrait, movies, videogram, instagramvideo   & \\
\hline
\multirow{2}{*}{1, 2, 7} & \multirow{2}{*}{\textsf{miscellanea}} & whpthroughthetrees, ig_captures, whpmyhometown, whpliquidlandscape,  whpemptyspaces, whpmotherlylove, & \multirow{2}{*}{16,573}  \\
&  &  whpthanksdad, whpstraightfacades, whpmyfavoriteplace, whpfirstphotoredo,  whpstrideby & \\
\hline
 8, 18   & \textsf{travel} & worldunion, whpmyfavoriteplace, travel, world\_shotz, worldcaptures, worldplaces, igworldclub & 1,200 \\
\hline
\multirow{2}{*}{3, 4, 5, 17, 11}  & \multirow{2}{*}{\textsf{attention-seeking}} & instagood, instamood, photooftheday, pleasecomment, pleaseshoutout, teamfollowback, & \multirow{2}{*}{5,794}\\
& & igers, picoftheday, instadaily, bestoftheday, webstagram, iphonesia, igdaily   & \\
\hline
\multirow{2}{*}{9, 15} & \multirow{2}{*}{\textsf{photo art}} &  whpsilhouettes, whpselfportrait, whplookingup, whpreflectagram, selfie, & \multirow{2}{*}{11,882}\\
&  & blackandwhite, whpbehindthelens, whpstilllife, silhouette, bnw, monochrome   & \\
\hline
\end{tabular}
}
\end{table*}

\subsection{Temporal trends and clustering}
\label{sec:Q5}

In our fifth research question  (\textbf{Q5}), we analyze how   lurking trends evolve, focusing on unveiling the structures hidden in such evolving trends. 

 We pursued this goal as a task of clustering of time series representing the users' lurking profiles.  
The basis for this clustering analysis lies in repeatedly applying our LurkerRank   to successive  snapshots of a network dataset. Since the  snapshots can vary in size, LurkerRank scores were first normalized to be comparable across different times.  
We then  generated a time series of the  normalized LurkerRank scores for every user in the dataset.  The resulting set of time series was the input for our clustering task.  

We adopted a \textit{soft clustering} approach to group the  time series of LurkerRank scores.  This implies that a time series is allowed to obtain fuzzy memberships to all clusters.  
Our choice is motivated by suspicion  that the natural clusters to be detected in this kind of time-course data could not be well-separated, rather they could be frequently overlap.    
A suitable method to detect clusters in this kind of data is based on \textit{fuzzy c-means clustering}. 
We used a particularly efficient implementation, provided by the Mfuzz R-package tool,\footnote{http://www.bioconductor.org/packages/release/bioc/html/Mfuzz.html.}    
based on minimization of the weighted square error function. 
Note that since the clustering is performed in Euclidean space, the time series  were standardized to have a mean value of zero and a standard deviation of one. This preprocessing step ensures that series with similar variations   are close in Euclidean space.    
As concerns the setting of the fuzzifier  and the number of clusters required by the clustering algorithm, we follow the methodology suggested in~\cite{SchwammleJ10}, and summarized 
in our submission support page  available at \textit{http://uweb.dimes.unical.it/tagarelli/timelr/}.  
 
Figure~\ref{fig:ts-clustering} shows some of the clustering results we obtained on the evaluation datasets. For this analysis, we initially selected  the top-25\% lurkers of the   snapshot at time zero, 
 then   kept only those users appearing in at least 50\% of the subsequent snapshots.  Results correspond to different scenarios, both in terms of time-granularity (which impacts on the time series length) and type of relation (i.e., comments, favorite-marks, likes plus comments) underlying the graphs from which the time series  were generated. 
Note that the membership values of time series are color-encoded in the plots, which    facilitates the identification of temporal patterns in the clusters.

It can be noted from the figure that some cases are characterized by quite evident trends. For instance, on \data{Flickr},  cluster\#2 groups lurkers whose behavior (lurking scores) evolves in the form of a series with an initial  plateau followed by an increasing ramp and then a decreasing ramp, finally by a new stagnation trend. 
Similar is the situation depicted by   cluster\#3 on \data{Flickr}. 
On \data{FriendFeed}, clusters\#1-\#3-\#4 present a  more or less marked period of roughly constant lurking behavior  between the 24th and 36th weeks, along with various peaks in the heads or tails  of the series, which would hint at particularly critical (passive) periods of lurking. 
In general,    more time-consuming actions (i.e., comments on \data{Instagram}, like+comments on \data{FriendFeed}) tend to correspond to trends that present sharper upward/downward shifts, and to clusters with more noisy data. 
Finally, note that except for    cluster\#1 on \data{Flickr}, lurking series do not tend to group into  decreasing trends, which would suggest that 
lurkers are not likely to spontaneously ``de-lurk'' themselves, i.e., to turn their behavior into a more active participation to the community life.

\subsection{Topical evolution}
\label{sec:Q6}

\begin{figure*}[t!]
\centering
\begin{tabular}{cccc} 
\includegraphics[width=0.22\textwidth]{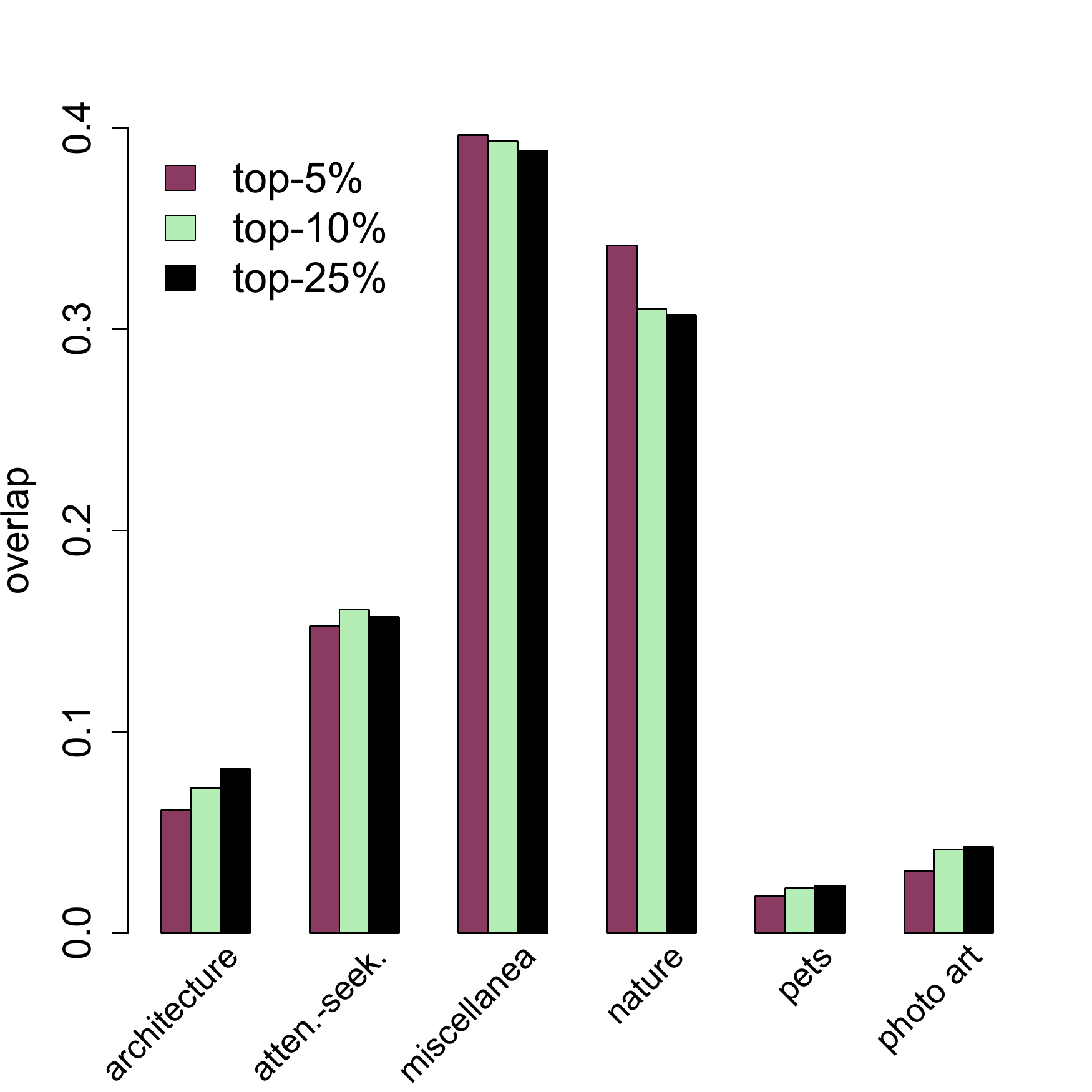} & 
\includegraphics[width=0.22\textwidth]{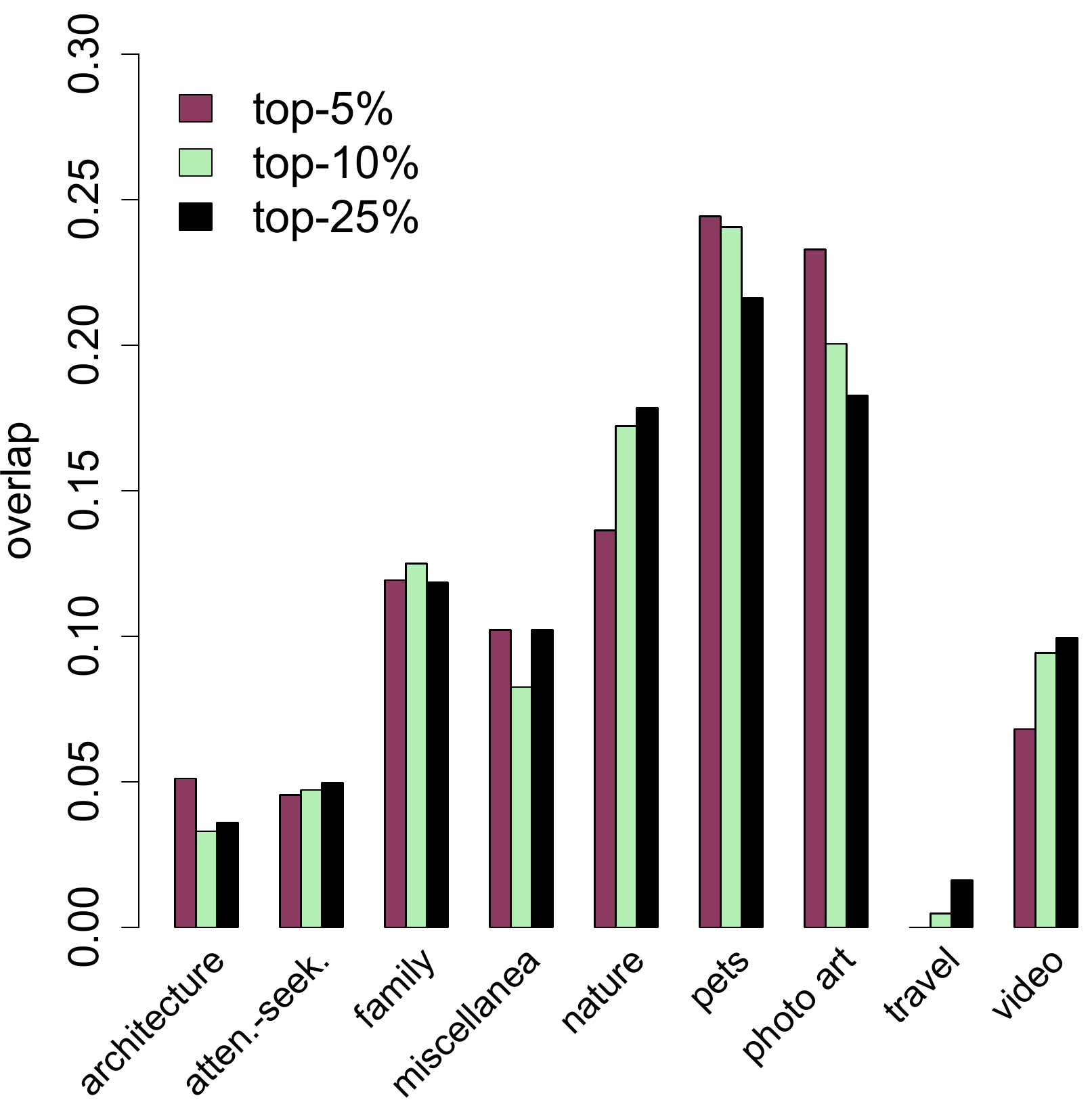} & 
\includegraphics[width=0.22\textwidth]{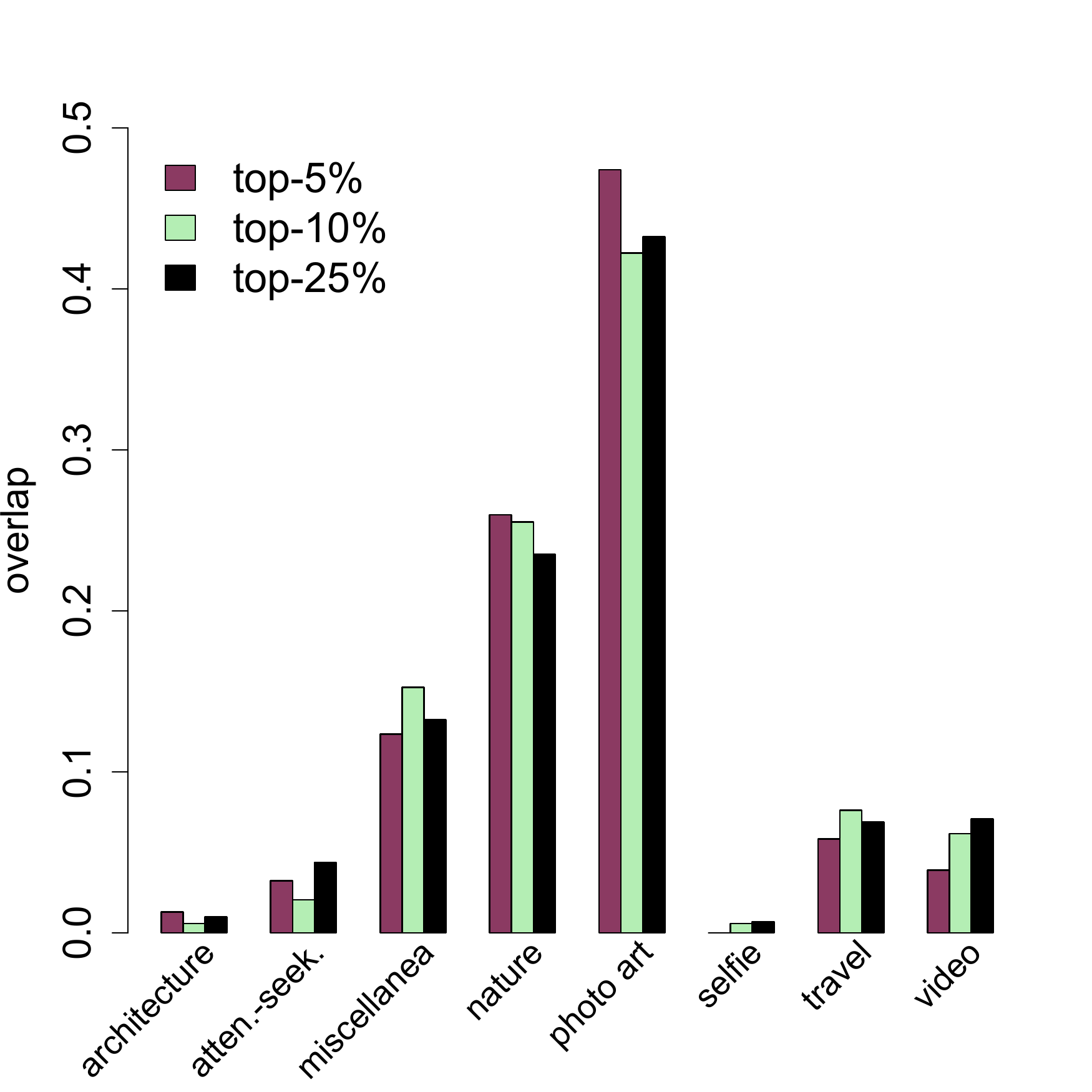} & 
\includegraphics[width=0.22\textwidth]{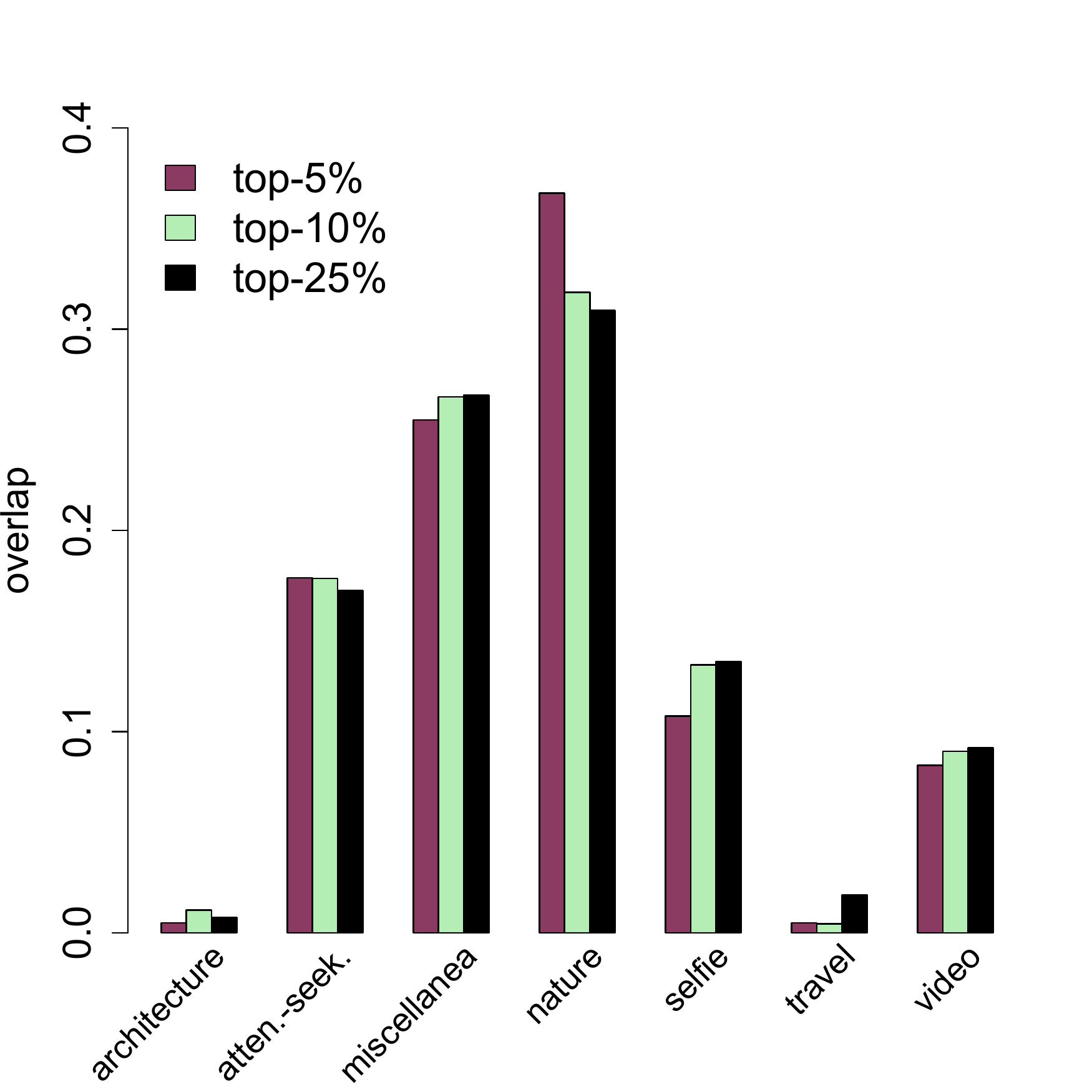} \\
(a) & (b) & (c) & (d) 
\end{tabular}
\caption{Overlap between the top-ranked lurkers detected on the snapshot graph and the top-ranked lurkers detected on each of the topic-specific subgraphs of the snapshot, at top-5\%, 20\%, and 25\%, over  the quarters of year 2013 in \data{Instagram} (first quarter on the left, last quarter on the right).}
\label{fig:topics-overlaps}
\end{figure*}

 Our six research question  (\textbf{Q6}) concerns the analysis of   topic-sensitive evolution patterns of lurking behavior. This  involves a characterization of the topical usage of lurkers, of how their topical patterns evolve and whether these may differ from those of the other users.  
 
To answer this question, we employed a  \textit{statistical topic model} to learn the topics of interest exhibited by the users.   More specifically, we used an efficient implementation provided in the \emph{gensim} library\footnote{http://radimrehurek.com/gensim/.}  of the well-known Latent Dirichlet Allocation (LDA)~\cite{BleiNJ03}. For the sake of brevity, we focus here on the presentation of results that we obtained on the \data{Instagram} dataset, for which we regarded all media of a user as a single document and the  tags assigned by users to their media as document features. We filtered out tags occurring in less than five documents or in more than 75\% of the documents in the collection. We tested our topic model with  5 to 50 latent topics, in increments of 5, executing up to 100 iterations; upon a manual inspection of the description of topics learned by the LDA models, we adopted the   model with 20 topics  as the most ``interpretable'' one. Our decision was indeed taken based on obtaining a topic description as sharp and rich as possible in terms of both characteristic and discriminating features.  We remark that   the topics extracted by our selected LDA model are consistent with a previous study on topical interests that was performed on a similar dump of the Instagram media dataset~\cite{HT14}. That study showed how the most popular tags in Instagram concern a limited number of categories, or coarse-grain topics, which include:   nature, travels, photography-related technical aspects, usage of popular applications for photo/video editing and publishing (e.g., Latergram, VSCO Cam), attention-seeking and microcommunity-focused tags (e.g., \#photooftheday, \#igmaster,  \#justgoshoot, \#iphonesia).

\begin{figure*}[t!]
\centering 
\includegraphics[width=0.65\textwidth]{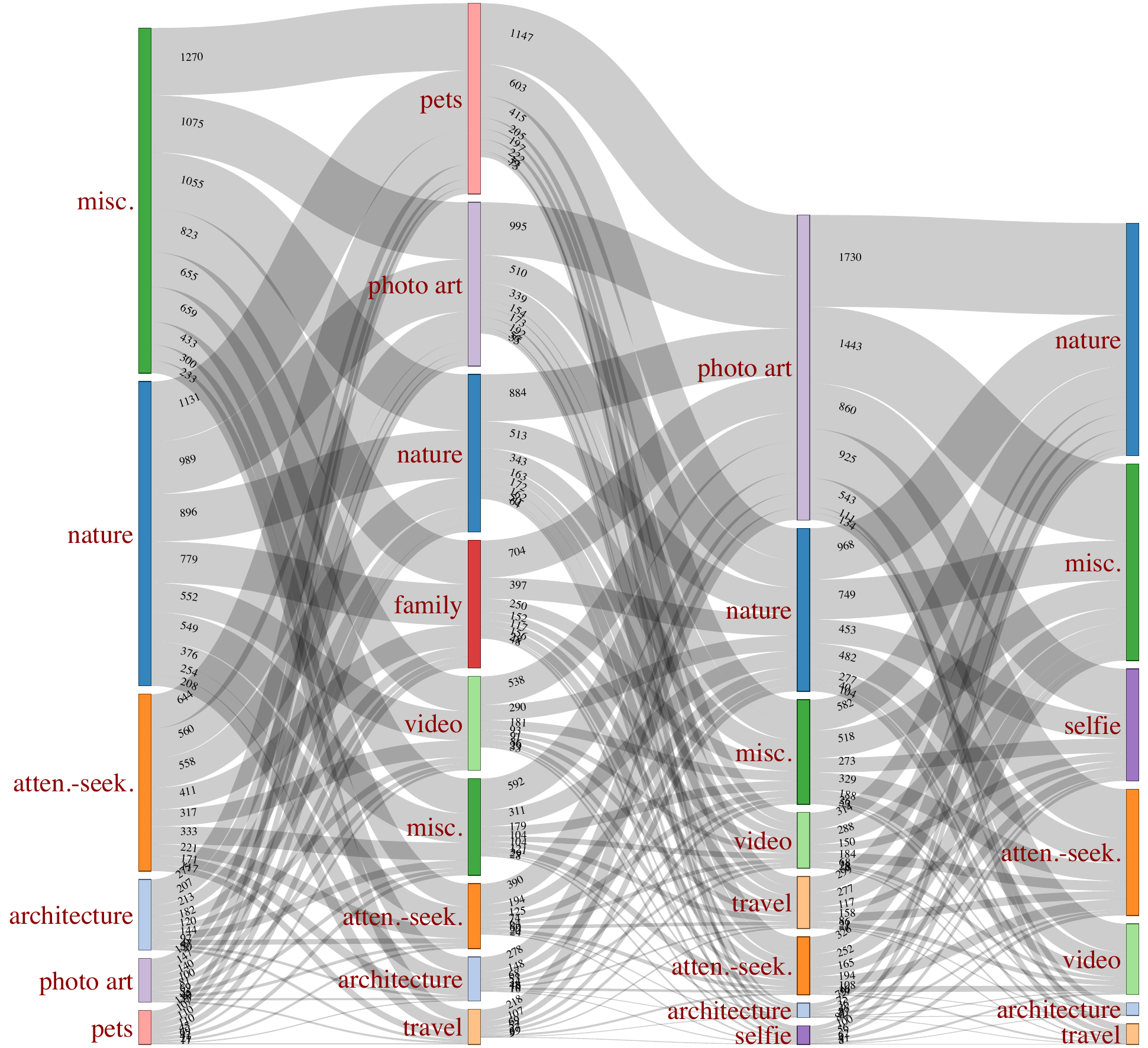} \\  \vspace{2mm}
\includegraphics[width=0.65\textwidth]{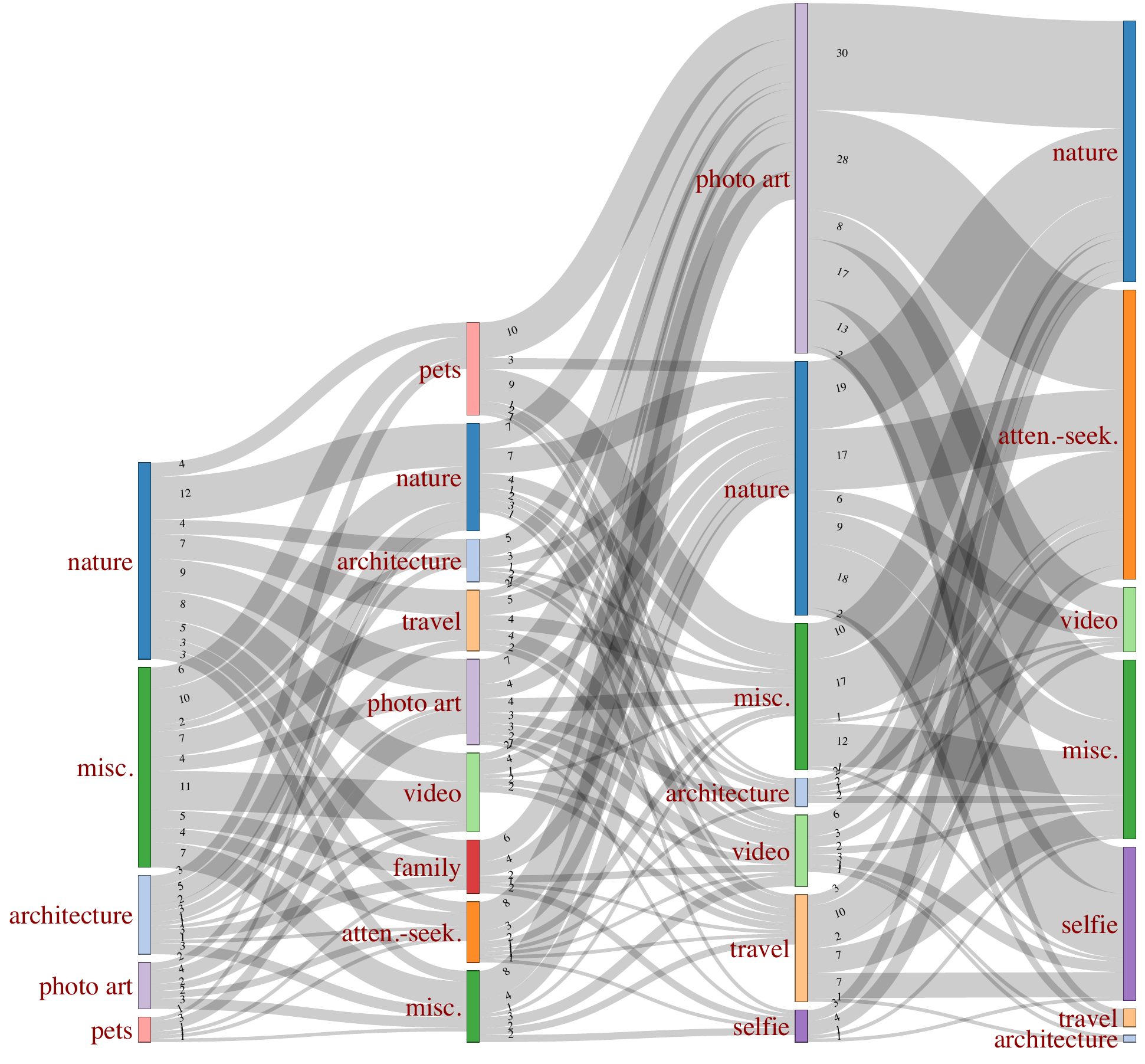}   
\caption{Topic evolution on  \data{Instagram}: all users (top) vs. top-25\%  lurkers (bottom). Levels correspond to quarters of year 2013. Each  vertical colored box represents a state as an aggregation of  topics, which are learned from the network contents at a given time (level).  Gray curves correspond to users transitioning from state to state. The portion of each state that does not have outgoing gray lines are users that end in this state. States are labeled with their description and their frequency, i.e., the number of users that are assigned to that topic at that level;  gray curves are proportional to the topic level frequencies. (Best viewed in color.)}
\label{fig:sankey}
\end{figure*}

We used the learned 20-topic LDA model to induce  topic-sensitive subgraphs from the \data{Instagram} user network. 
To derive each of these subgraphs, we first aggregated the finer-grain topics learned by LDA  into thematically-cohesive \textit{topic-sets}, then  every  user  was assigned to the topic-set that maximizes the likelihood in the LDA per-document topic distributions.     
Table~\ref{tab:topics} shows a (partial) description of the topics learned by LDA, along with the chosen labels for  the derived topic-sets  and the impact on the size (number of nodes) of the induced  topic-specific subgraphs. 
Note that we also include the \textsf{miscellanea} topic-set which covered  all user documents whose LDA topic distributions were characterized by a quite high topical entropy---again, this is in accord with our study in \cite{HT14}, which highlighted that most users adopt few tags to annotate their media, but also that  popular users have higher topical entropy values (i.e., topic specialization is not relevant).

Upon the extraction of  topic-specific subgraphs, we looked for clues about major topics (i.e., frequently used tags) that  characterize lurkers. To do this, we compared  the top-ranked lurkers detected in the full, topic-independent graph and the top-ranked lurkers detected in each of the topic-specific subgraphs, 
for a given fraction of top-ranked lurkers (varying at 5\%, 10\% and 25\%). More precisely,  for each topic, we computed an overlap score  as the intersection between the set of top-ranked lurkers   in the topic-specific subgraph and the   set of top-ranked lurkers in the full graph,  divided by the sum of intersection values obtained over all topics.  
Results (not shown) put in evidence  a relatively good matching  between the top-ranked lurkers in the full graph and those relating to the subgraph specific of the \textsf{photo art} topics (overlap ranging from 0.37 at top-5\% to 0.25 at top-25\%), followed by \textsf{nature} (overlap around 0.13-0.14) and \textsf{attention-seeking} tag topics (overlap of 0.13-0.10). Other tags specific to any other topic-set in Table~\ref{tab:topics} correspond to low overlaps (below 0.05), with the exception of \textsf{miscellanea} whose corresponding overlaps vary from 0.28 to 0.38 by increasing the size of top-ranked lurkers under consideration.    
 More interestingly, we repeated the above evaluation over selected temporal snapshots of the \data{Instagram} network.  
 Figure~\ref{fig:topics-overlaps} shows results obtained over the quarters of year 2013, which corresponds to the timespan that covers most user actions and interactions in our \data{Instagram} dataset. As can be seen from the plots in the figure, the topic usage behavior of lurkers in each snapshot is mainly characterized by tags that belong to one or more topic-sets; particularly, \textsf{photo art} in the third and second quarter, \textsf{nature} in the last quarter but also in the other ones,  \textsf{pets} in the second quarter. It is interesting to observe that, with the exception of the first quarter snapshot,  \textsf{miscellanea} tags are not  a frequent choice of lurkers,  i.e., lurkers are more likely to  focus on contents (media) that are well categorized into only one of the identified topic-sets.

We further analyzed the evolution of topic interests over time. In this regard, we hypothesized that lurkers might exhibit   patterns of  topical interests that do not significantly differ from those of the other (active) users.   
Figure~\ref{fig:sankey} shows two transition diagrams which offer a view of how the topical usage  patterns change from one state (i.e., topic-set) to another, over  the quarters of year 2013 in \data{Instagram},  for  all users as well as for all top-25\% lurkers. 
Let us first consider  the topical evolution of all users (top of Fig.~\ref{fig:sankey}).  Here we observe that the various levels (i.e., quarters of  year 2013) are characterized by a core of topic-sets which,  although with varying proportions, are always present over time (i.e., \textsf{nature}, \textsf{attention-seeking}, \textsf{architecture}, and \textsf{miscellanea}). Other topic-sets (e.g., \textsf{pets} and \textsf{photo art}) may correspond to temporary  interests of users, as they are present only in some of the levels.   
Topical usage patterns of the users tend to continuously change over time. We in fact observe transitions  from one topic-set state in a level to each of the other states in the next level. Note that such a high dynamicity is not surprising, which is explained by the inherent softness of topic categorization underlying the tags used for the uploaded media in Instagram and similar OSNs; in other terms,  users can often adopt tags that naturally belong to more than one topic-set to annotate their media, according to the type of photo or video (e.g., a skyline photo can be equally relevant to the categories \textsf{photo art}, \textsf{travel}, \textsf{attention-seeking}). 
However, as it happens at the second level (quarter),   all topic-set states can also show a moderate stability, since a fraction of users (about 20\%) do not transition out of a topic-set state once they enter it.  

Topical usage transitions in the graph of the top-ranked lurkers (bottom of  Fig.~\ref{fig:sankey}) are also highly dynamic. The topic-sets per level are either the same as or a subset of those in the all-users graph, showing different relative proportions (i.e., frequency of usage)  in some cases (e.g.,  \textsf{family} in the second level, \textsf{attention-seeking} in the fourth level). 
This would hence confirm our initial hypothesis  that lurkers tend to  show patterns of  topical interests that do not significantly differ from the ones of all users. 
A major difference with the all-users graph however is that  in some cases more transitions flow out from a topic-set state than the incoming ones, which corresponds to the behavior of lurkers as ``newcomers'', i.e., lurkers that were not present in the immediately preceding  snapshot graph, but could be in earlier snapshots (cf. Section~\ref{sec:Q2}). For instance, while several lurkers showing different interests at the second level end in the \textsf{photo art} state at the third level,  a nearly equal proportion of new lurkers   start from that state, then transition towards different topic-sets.

\subsection{Time-aware ranking of lurkers}
\label{sec:Q7}

Our final research question ({\bf Q7}) is devoted to the analysis of time-aware ranking of lurkers. We assess the presumed benefits derived from the use of  our proposed  time-transient and    time-cumulative ranking models on the quality of lurker ranking solutions.  
In the following, we first present our evaluation methodology, then we discuss   effectiveness results obtained by our time-aware LurkerRank methods and competing methods.

 \subsubsection{Data-driven evaluation} 
 \label{sec:ddrank}
Evaluating lurking in OSNs is a hard problem  to deal with, because of  the lack of ground-truth data  for lurker ranking. 
In the attempt of simulating  a ground-truth evaluation, 
we build on top  of our previous studies~\cite{ASONAM,SNAM14}. We  generate a  \textit{data-driven ranking} (henceforth \algo{DD})   
for every network graph and use  it to assess  the proposed and competing methods. 
However, in contrast to the data-driven rankings defined in~\cite{ASONAM,SNAM14}, 
here we focus on the amount of \textit{actions} and \textit{interactions} users perform over a   time interval. Formally, given   $v \in \V$ and time interval $T$, the data-driven ranking score assigned to $v$ at time $T$ is computed as:
\begin{equation}\label{eq:refrank}
r^{*}_T(v) = \frac{   \sum_{u \in B_v} \sum_{t \in T} n_t(u,v)    } {    \sum_{t \in T} n_t(v)    }
\end{equation}
where $n_t(v)$ denotes the number of actions that  $v$ performed  at time $t$ to create new contents (e.g., media uploads), and $n_t(u,v)$  denotes the number of information-consumption actions at time $t$ performed by $v$ in response to a specific post by $u$. 
Given the characteristics of our selected  datasets, we compute $n_t(u,v)$ as the number of  ``favorite''  or ``like'' actions by $v$ in relation to a media posted by $u$.    
We observe however that, in general, an information-consumption action does not necessarily imply that the user  will produce  \textit{visible} information such as   posting a   ``like'' or  ``comment'' in response to another user's post.      
Within this view, timestamped information-consumption actions could  refer   to the \textit{latent} or \textit{silent} interactions, i.e., the actions of reading or watching   produced contents; unfortunately, 
it is not easy to build OSN datasets that are resource-rich in terms of latent interactions, mainly due to  privacy policies and API limitations currently imposed by all main OSN services. We will leave the opportunity of evaluating the lurker ranking problem focusing  on latent interactions as a future work (cf. Section~\ref{sec:challenges}).

 \subsubsection{Competing methods and assessment criteria}   
We compared our proposed methods \algo{Ts-LR} and \algo{Te-LR} 
against the  early (i.e., time-unaware) LurkerRank (\algo{LR})~\cite{SNAM14,ASONAM}.   
Note that applying \algo{LR} on interaction graphs is here assumed to be consistent with its definition:    an interaction graph is a subset of the followship graph, however  provided that only visible interactions are taken into account, which is indeed our evaluation setting.   
 
As we presented in Section~\ref{sec:TLR}, our proposed time-aware LurkerRank methods are defined upon   two models of temporal graph. Therefore, we carried out   \algo{Ts-LR} and \algo{Te-LR} on different types of  graphs, dubbed  transient  and  cumulative  snapshots, respectively.  
More precisely, we hereinafter refer to  a  \textit{monthly, transient snapshot} as a snapshot whose timespan is   28 days (cf. Section~\ref{sec:data-description}). We also use the term  \textit{monthly, cumulative snapshot} to denote a snapshot   covering   a time window that is one month larger than the previous snapshot; moreover, the start time is fixed for all monthly, cumulative snapshots considered on a network dataset, thus the size of cumulative snapshots follows a non-decreasing function. 

We also included in the evaluation the   \algo{T-Rank} algorithm~\cite{BerberichVW05}, which is a time-aware adaptation of PageRank; as we   discuss more in detail in Section~\ref{sec:relatedwork}, \algo{T-Rank} was chosen as a competitor since, like our proposed methods,  it also embeds   notions   of freshness and activity.  
We used the setting of the parameters in \algo{T-Rank}  as suggested in~\cite{BerberichVW05}, using uniform values for both the types of coefficients that  control the use of temporal information in the time-aware method: the four $w_{si}$ coefficients used to determine the random jump probabilities, and the six $w_{ti}$ coefficients that  give the transition probabilities of the random surfer. 
Note also that \algo{T-Rank}  was  involved  only in our transient evaluation case, since the algorithm was designed to work in transient snapshots. 
In this regard, we evaluated  \algo{T-Rank} on monthly,  transient snapshots, setting the temporal window of interest to the last week of the month, and the tolerance interval to the first three weeks. This choice was made in order to give more importance to \textit{recent} temporal information (w.r.t. the end-time of each target snapshot).

To evaluate  the ranking performance of the various methods, 
we used two  well-known assessment criteria in ranking tasks, namely 
\emph{Kendall-tau rank correlation coefficient}~\cite{abdi07}, and 
\emph{Fagin's intersection metric}~\cite{FaginKS03}.  
Kendall-tau correlation evaluates the similarity between two rankings, expressed as sets of ordered pairs,  based on the number of inversions of   pairs which are needed to transform one ranking into the other: 
\begin{equation*}
\tau(\L',\L'') = 1 - \frac{2\Delta(\mathcal{P}(\L'),\mathcal{P}(\L''))}{M(M-1)}
\end{equation*}
Above, $\L'$ and $\L''$ are the two rankings to be compared, $M=|\L'|=|\L''|$ and $\Delta(\mathcal{P}(\L'),\mathcal{P}(\L''))$ is the symmetric difference distance between the two rankings, calculated as number of unshared  pairs between the two lists.    
The score returned by $\tau$ is in the interval $[-1,1]$, where a value of $1$ means that the two rankings are identical and a value of $-1$ means that one ranking is the reverse of the other.    
Fagin measure allows for determining how well two ranking lists are in agreement with each other, taking into account top-weightedness and partial rankings.    
Applied to any two top-$k$ lists $\L',\L''$, the Fagin score is defined as:  
\begin{equation*} \label{eq:fagin}
F(\L',\L'',k) = \frac{1}{k}\sum_{q=1}^k \frac{|\L'_{:q} \cap \L''_{:q}|}{q}
\end{equation*}
where $\L_{:q}$  denotes the sets of nodes  from the 1st  to the $q$th position in the   ranking.
Therefore,  $F$ is the average over the sum of the weighted overlaps based on the first $k$ nodes  in both rankings;  experimental results we shall present  in  Section~\ref{sec:results} correspond to $k$ fixed to the 25\% of the ranking lists being compared (denoted as $F$@25\%).  
Note that the $F$ score is in the interval $[0,1]$, where  $1$ means total agreement and $0$ means total disagreement.

 \subsubsection{Results}
 \label{sec:results}
We focus our evaluation of time-aware LurkerRank methods on the \data{Flickr} and \data{Instagram} datasets. A   major reason  for this choice  is that  we wanted to evaluate our proposed  methods against snapshots extracted from  a social (followship) network as well as from an interaction network. The former scenario was evaluated thanks to the   timestamped   followship information that is only available in \data{Flickr}. For the latter scenario we used the timestamped information about favorite-markings available from \data{Flickr},  and the timestamped information about comments available from \data{Instagram};  note that this allowed us to evaluate the performance of the methods over   snapshot graphs that correspond to two types of interactions, i.e., either ``likes'' or comments.  
Note also that we left \data{FriendFeed} out of consideration  because it is less rich than \data{Instagram} in terms of timestamped information about comments; moreover, comments in \data{FriendFeed} concern user interactions corresponding to a relatively smaller portion of social graph than in \data{Instagram} (cf. Section~\ref{sec:data-description}).

\begin{table}[t!]
\caption{Kendall-tau correlation and Fagin's intersection performance w.r.t.  \algo{DD} on \textit{monthly, cumulative snapshots} of the \data{Flickr} \textit{followship} network:  comparison between \algo{Te-LR} against  \algo{LR}. (Bold values correspond to the best performance per assessment criterion.)}
\centering 
\scalebox{0.85}{
\begin{tabular}{|l||c|c||c|c|}
\hline
\textit{snapshot}  &\multicolumn{2}{|c||}{$\tau$}   &  \multicolumn{2}{|c|}{$F$@25\%} \\
\cline{2-5}
 \textit{(end time)}             &  \algo{Te-LR} & \algo{LR}  
         &  \algo{Te-LR} & \algo{LR} \\
\hline \hline
2006-11-30	& 0.145 & 0.021 & 0.138 & 0.135 \\    
2006-12-28	& 0.148 & 0.007 & 0.142 & 0.135 \\
2007-01-25	& 0.159 & -0.005 & 0.152 & 0.138 \\
2007-02-22	& 0.178 & -0.017 & 0.156 & 0.135 \\
2007-03-22	& \HL{0.186} & -0.019 & \HL{0.162} & 0.133 \\
2007-04-19	& \HL{0.186} & -0.017 & 0.161 & 0.133 \\
2007-05-17	& \HL{0.186} & -0.015 & 0.161 & 0.133 \\
 \hline
\end{tabular}
}
\label{tab:Flickr-followship}
\end{table}

Table~\ref{tab:Flickr-followship} shows performance results obtained  on month\-ly  snapshots of the \data{Flickr} followship network. Here we left  \algo{Ts-LR} out of consideration since cumulative snapshots   make more sense than   transient snapshots when followship relations are taken into account. In other terms, since the set of  followships  in a network grows progressively (unless unfollowing actions are permitted), it is  unfair to  consider transient snapshots which would ignore all the relations created before the selected time windows.     
From the table, we observe that \algo{Te-LR}  always obtains a higher correlation with \algo{DD} than \algo{LR}, with gains in terms of Kendall-tau ranging from   $0.124$ to  $0.205$, and gains in terms of Fagin's intersection up to  
$0.029$. It should be noted that even though \algo{LR} shows  negative Kendall-tau correlation w.r.t. \algo{DD} for more recent (i.e., cumulatively aggregated)  snapshots, Fagin's intersection values are not so distant   from the ones obtained by \algo{Te-LR}.  This would indicate that  \algo{Te-LR} corresponds to a superior lurker-ranking model  when applied to  all users in the network as well as only to the most prominent ones as lurkers in the network,  while the latter would be suboptimally detected by \algo{LR}.

\begin{table}[t!]
\caption{Kendall-tau correlation and Fagin's intersection performance w.r.t.  \algo{DD} on \textit{monthly, transient snapshots} of the \data{Flickr} \textit{interaction} network:  comparison between  \algo{Ts-LR} against  \algo{LR} and \algo{T-Rank}. (Bold values correspond to the best performance per assessment criterion.)}
\centering 
\scalebox{0.85}{
\begin{tabular}{|l||c|c|c||c|c|c|}
\hline
\textit{snapshot}  &\multicolumn{3}{|c||}{$\tau$}   &  \multicolumn{3}{|c|}{$F$@25\%} \\
\cline{2-7}
 \textit{(end time)}             &  \algo{Ts-LR} & \algo{LR} &  \algo{T-Rank} 
         &  \algo{Ts-LR} & \algo{LR} &  \algo{T-Rank} \\
\hline \hline
2006-10-05 & 0.156 & -0.014 & -0.154 & 0.165 & 0.138 & 0.069 \\
2006-11-02 & \HL{0.169} & -0.023 & -0.167 & 0.174 & 0.133 & 0.068 \\
2006-11-30 & 0.167 & -0.022 & -0.159 & 0.166 & 0.138 & 0.063 \\
2006-12-28 & 0.155 & -0.003 & -0.147 & 0.171 & 0.133 & 0.068 \\
2007-01-25 & \HL{0.169} & -0.025 & -0.16 & 0.167 & 0.134 & 0.073 \\
2007-02-22 & 0.167 & -0.018 & -0.159 & \HL{0.178} & 0.14 & 0.066 \\
2007-03-22 & 0.164 & 0 & -0.14 & 0.158 & 0.143 & 0.073 \\
\hline
\textit{avg.}  &   \textit{0.164}	 &  \textit{-0.015}	&   \textit{-0.155}	&   \textit{0.168}	  &  \textit{0.137}  &	\textit{0.069} \\
 \hline
\end{tabular}
} 
\label{tab:Flickr-interaction-transient}
\end{table}

\begin{table}[t!]
\caption{Kendall-tau correlation and Fagin's intersection performance w.r.t.  \algo{DD} on \textit{monthly, cumulative snapshots} of the \data{Flickr} \textit{interaction} network:  comparison between  \algo{Te-LR} against  \algo{LR}. (Bold values correspond to the best performance per assessment criterion.)}
\centering 
\scalebox{0.85}{
\begin{tabular}{|l||c|c||c|c|}
\hline
\textit{snapshot}  &\multicolumn{2}{|c||}{$\tau$}   &  \multicolumn{2}{|c|}{$F$@25\%} \\
\cline{2-5}
\textit{(end time)}              &  \algo{Te-LR} & \algo{LR} 
         &  \algo{Te-LR} & \algo{LR} \\
\hline \hline
2006-10-05 & 0.156 & -0.014 & 0.165 & 0.138 \\
2006-11-02 & 0.177 & -0.024 & 0.175 & 0.141 \\
2006-11-30 & 0.179 & -0.030 & 0.173 & 0.140 \\
2006-12-28 & 0.185 & -0.033 & \HL{0.179} & 0.141 \\
2007-01-25 & 0.191 & -0.046 & 0.177 & 0.141 \\
2007-02-22 & 0.197 & -0.050 & 0.175 & 0.139 \\
2007-03-22 & \HL{0.200} & -0.050 & 0.176 & 0.138 \\
 \hline
\end{tabular}
} 
\label{tab:Flickr-interaction-cumulative}
\end{table}

Table~\ref{tab:Flickr-interaction-transient} and Table~\ref{tab:Flickr-interaction-cumulative} still focus on \data{Flickr}, however on a different scenario in which the ranking methods are applied over snapshots of the \data{Flickr} interaction network.  
Results from  Table~\ref{tab:Flickr-interaction-transient}, which correspond to transient snapshots, show the better performance of  \algo{Ts-LR}  against   \algo{LR} and \algo{T-Rank},   according to both assessment   criteria (with average Kendall-tau correlation of   $0.164$ and average Fagin's intersection   of $0.168$). In particular,   \algo{Ts-LR} outperforms  \algo{T-Rank}, with an average gain of $0.319$ Kendall-tau  of $0.100$  Fagin's intersection. Also,   \algo{Ts-LR} achieves higher Kendall-tau correlation w.r.t. \algo{DD} than   \algo{LR} (up to $0.194$, with an average gain of $0.179$), while the difference in terms of  Fagin's intersection is smaller. 
While Kendall-tau values obtained by \algo{LR} and \algo{T-Rank} are  negative, it should be noted that \algo{T-Rank} also shows very low Fagin's intersection values (always below  $0.1$) 
while \algo{LR} maintains  a certain  intersection with the top of \algo{DD} (average Fagin's intersection  of $0.137$).
Analogous conclusions can be drawn from the evaluation of \algo{Te-LR} and  \algo{LR} over monthly, cumulative snapshots of the \data{Flickr} interaction network (Table~\ref{tab:Flickr-interaction-cumulative}). Again, a significant difference in terms of  Kendall-tau correlation is observed between the performance of \algo{Te-LR} and   \algo{LR} (average gap of $0.250$, with \algo{LR} always showing negative correlation w.r.t.  \algo{DD}), while Fagin's intersection values  of \algo{LR} is relatively lower than the ones obtained by  \algo{Te-LR} (average gain of only $0.038$ in favor of \algo{Te-LR}).

\begin{table}[t!]
\caption{Kendall-tau correlation and Fagin's intersection performance w.r.t.  \algo{DD} on \textit{monthly, transient snapshots} of the \data{Instagram} \textit{interaction} network:  comparison between  \algo{Ts-LR} against  \algo{LR} and \algo{T-Rank}. (Bold values correspond to the best performance per assessment criterion.)}
\centering 
\scalebox{0.85}{
\begin{tabular}{|l||c|c|c||c|c|c|}
\hline
\textit{snapshot}  &\multicolumn{3}{|c||}{$\tau$}   &  \multicolumn{3}{|c|}{$F$@25\%} \\
\cline{2-7}
  \textit{(end time)}            &  \algo{Ts-LR} & \algo{LR} &  \algo{T-Rank} 
         &  \algo{Ts-LR} & \algo{LR} &  \algo{T-Rank} \\
\hline \hline
2012-07-04 & \HL{0.367} & 0.145 & 0.235 & 0.227 & 0.170 & 0.192 \\
2012-08-01 & 0.120 & 0.107 & 0.179 & 0.244 & 0.221 & 0.103 \\
2012-08-29 & 0.197 & 0.153 & 0.140 & \HL{0.270} & 0.234 & 0.202 \\
2012-09-26 & 0.255 & 0.211 & 0.111 & 0.246 & 0.205 & 0.120 \\
2012-10-24 & 0.200 & 0.166 & 0.126 & 0.237 & 0.205 & 0.148 \\
2012-11-21 & 0.254 & 0.234 & 0.137 & 0.185 & 0.154 & 0.108 \\
2012-12-19 & 0.231 & 0.201 & 0.119 & 0.230 & 0.201 & 0.180 \\
2013-01-16 & 0.236 & 0.211 & 0.095 & 0.257 & 0.235 & 0.126 \\
2013-02-13 & 0.253 & 0.221 & 0.110 & 0.234 & 0.210 & 0.155 \\
2013-03-13 & 0.203 & 0.166 & 0.179 & 0.195 & 0.180 & 0.154 \\
2013-04-10 & 0.249 & 0.225 & 0.137 & 0.190 & 0.188 & 0.167 \\
2013-05-08 & 0.282 & 0.253 & 0.099 & 0.249 & 0.235 & 0.144 \\
2013-06-05 & 0.282 & 0.256 & 0.139 & 0.219 & 0.214 & 0.159 \\
2013-07-03 & 0.247 & 0.216 & 0.136 & 0.227 & 0.211 & 0.135 \\
2013-07-31 & 0.218 & 0.201 & 0.157 & 0.191 & 0.190 & 0.131 \\
2013-08-28 & 0.236 & 0.207 & 0.143 & 0.201 & 0.181 & 0.165 \\
2013-09-25 & 0.268 & 0.248 & 0.132 & 0.218 & 0.202 & 0.130 \\
2013-10-23 & 0.209 & 0.191 & 0.103 & 0.183 & 0.173 & 0.156 \\
2013-11-20 & 0.234 & 0.217 & 0.093 & 0.231 & 0.216 & 0.143 \\
2013-12-18 & 0.226 & 0.211 & 0.115 & 0.229 & 0.224 & 0.126 \\
\hline
\textit{avg.}  &   \textit{0.238}	&  \textit{0.202}	&   \textit{0.134}	&   \textit{0.223}	&   \textit{0.203}	 &  \textit{0.147} \\
 \hline
\end{tabular}
}
\label{tab:Instagram-transient}
\end{table}

\begin{table}[t!]
\caption{Kendall-tau correlation and Fagin's intersection performance w.r.t.  \algo{DD} on \textit{monthly, cumulative snapshots} of the \data{Instagram} \textit{interaction} network:  comparison between  \algo{Te-LR} against  \algo{LR} and \algo{T-Rank}. (Bold values correspond to the best performance per assessment criterion.)}
\centering 
\scalebox{0.85}{
\begin{tabular}{|l||c|c|c||c|c|c|}
\hline
\textit{snapshot}  &\multicolumn{2}{|c||}{$\tau$}   &  \multicolumn{2}{|c|}{$F$@25\%} \\
\cline{2-5}
\textit{(end time)}       &  \algo{Te-LR} & \algo{LR}  
         &  \algo{Te-LR} & \algo{LR} \\
\hline \hline
2012-07-04 & \HL{0.366} & 0.145 & \HL{0.232} & 0.170 \\
2012-08-01 & 0.235 & 0.112 & \HL{0.232} & 0.203 \\
2012-08-29 & 0.239 & 0.074 & 0.210 & 0.123 \\
2012-09-26 & 0.233 & 0.090 & 0.185 & 0.116 \\
2012-10-24 & 0.211 & 0.097 & 0.187 & 0.127 \\
2012-11-21 & 0.203 & 0.101 & 0.180 & 0.128 \\
2012-12-19 & 0.188 & 0.094 & 0.173 & 0.131 \\
2013-01-16 & 0.175 & 0.092 & 0.178 & 0.146 \\
2013-02-13 & 0.160 & 0.088 & 0.174 & 0.148 \\
2013-03-13 & 0.148 & 0.079 & 0.166 & 0.144 \\
2013-04-10 & 0.143 & 0.079 & 0.166 & 0.152 \\
2013-05-08 & 0.137 & 0.078 & 0.163 & 0.155 \\
2013-06-05 & 0.126 & 0.072 & 0.163 & 0.156 \\
2013-07-03 & 0.123 & 0.076 & 0.163 & 0.159 \\
2013-07-31 & 0.115 & 0.073 & 0.163 & 0.162 \\
2013-08-28 & 0.112 & 0.077 & 0.166 & 0.166 \\
2013-09-25 & 0.105 & 0.075 & 0.165 & 0.167 \\
2013-10-23 & 0.098 & 0.075 & 0.171 & 0.177 \\
2013-11-20 & 0.091 & 0.075 & 0.179 & 0.187 \\
2013-12-18 & 0.088 & 0.079 & 0.182 & 0.191 \\
 \hline
\end{tabular}
}
\label{tab:Instagram-cumulative}
\end{table}

Results over the \data{Instagram} monthly snapshots are reported in Table~\ref{tab:Instagram-transient} and Table~\ref{tab:Instagram-cumulative}. 
Again, \algo{Ts-LR} and \algo{Te-LR} always perform  better than competitors, in the corresponding evaluation scenarios. Moreover, compared to the results obtained on \data{Flickr},   the performance of \algo{LR} is generally closer to that of the time-aware LurkerRank algorithms.  
In the transient evaluation case  (Table~\ref{tab:Instagram-transient}), Kendall-tau correlation is positive for all methods, with \algo{Ts-LR} showing higher correlation w.r.t.  \algo{DD} than \algo{T-Rank} (average gain of $0.104$), and similar correlation values when compared to \algo{LR} (average gain of $0.036$). An analogous situation can be depicted when considering Fagin's intersection scores, with gains  up to $0.141$ w.r.t.  \algo{T-Rank} and up to $0.057$ w.r.t. \algo{LR}. 
Differences between \algo{Te-LR} and \algo{LR} are even smaller when looking at the cumulative case (Table~\ref{tab:Instagram-cumulative}), with always decreasing Kendall-tau gains which range from the $0.221$ of the first snapshot, to the $0.01$ of the last one. Fagin's intersection values   are very similar in most cases (maximum gain of $0.087$), with \algo{LR}   performing comparably to or   better than  \algo{Te-LR} in the last snapshots.  
An explanation might be found according to a fact that we already observed in  Section~\ref{sec:Q1}, that is,  lurkers are more prone to perform actions like ``favorites'' or ``likes'' than to comment posts. Therefore,  the ``favorite/like'' type of interaction  would act  as a better discriminant than ``comment''  in capturing the lurker dynamics via a time-varying graph model.

 As a final remark, we observe  that \algo{Te-LR} has different overall behavior in \data{Flickr} and in \data{Instagram} time-evolving graphs, which corresponds  to two diverse time\-spans, i.e., 7 months in \data{Flickr} against  20 months in \data{Instagram}. In particular, for more recent snapshots, ranking performance of \algo{Te-LR} is generally increasing in \data{Flickr} and decreasing in \data{Instagram}, especially in terms of Kendall-tau.  This would suggest a certain sensitivity of  \algo{Te-LR} to long timespans, which might negatively affect the \algo{Te-LR} performance, even yielding worse results than the basic (i.e., time-unaware) LurkerRank.

\section{Related Work} 
\label{sec:relatedwork} 

\subsection{Lurking in social networks} 
 
 Research studies in social science and human-computer interaction have scrutinized the various definitions of lurking, analyzed the motivational factors for lurking, and devised the main strategies for de-lurking. For instance, Soroka and Rafaeli~\cite{SorokaR06}    investigated relations between lurking and cultural capital, i.e., a member's level of community-oriented knowledge, while 
       lurking was conceptualized in~\cite{CranefieldYH11,Muller12} in terms of the users' boundary spanning and knowledge brokering activities across multiple community engagement spaces.  
The implications behind  lurking were  also  analyzed from a group  learning~\cite{ChenC11},    peripheral participation~\cite{HalfakerKT13}, and   epistemological~\cite{SchneiderKJ13} perspectives.  
  
 Understanding lurkers in OSNs  refers to a scenario that has remained quite unexplored in computer science until recently.   
 Fazeen et al.~\cite{FazeenDG11}  addressed classification of  the various  actors in an OSN (i.e.,  leaders, spammers, associates, and lurkers). However, in that work the lurking problem is  treated margin\-ally, and in fact lurking cases are left out of experimental evaluation. Similarly,  Lang and Wu~\cite{LangW13} analyzed various factors  that influence lifetime of  users, 
also distinguishing between active and passive lifetime. In this regard, a number of  features is suggested to  promote usage among members of OSNs like Twitter and Buzznet. Moreover, while examining to what extent active and passive lifetime are correlated, the authors observed that the study of   passive lifetime requires to know   the user's last login date,  which is however unavailable for many OSNs including Twitter.   
Besides our   work on lurker detection and ranking~\cite{SNAM14,ASONAM} (previously discussed in Section~\ref{sec:backgroundLR}),   
in~\cite{ASONAM14} we started investigating how lurker behaviors    change over time. In that work we provided   a preliminary characterization of the lurking dynamics in terms of four out of the seven research questions that we have addressed in  Section~\ref{sec:analysis}.

\subsection{Time-aware PageRank} 
\label{sec:taPR}
Given the popularity of PageRank among researchers  in web searching of authoritative sources,  most solutions for time-aware ranking have been developed in the past years by resorting to PageRank-style  methods. 
Here we briefly discuss some of the most relevant studies,   which share   intuitions with our approach. 

One of the earliest methods that leverage the temporal dimension in authority ranking is TimedPageRank~\cite{YuLL04}. This is basically a weighted PageRank applied to a citation network in which 
the strength of every edge (citation) is weighted by an exponential decay function of  the citation   age. An aging-factor is also introduced to linearly penalize the scores w.r.t. the age of a publication. 
TimedPageRank   adopts a   graph model that represents a static picture of the network at a given time.  
By contrast, EventRank~\cite{Smyth05} takes into account the sequence of events and can track changes in ranking over time. Originally developed for a  (email) communication network, EventRank utilizes  potential flow to model the exchange of messages in a cumulative ranking fashion.   As mentioned in the Introduction, 
our time-aware lurker ranking methods also handles either approach (i.e., static or cumulative) to ranking.
 
Our proposed time-aware lurker ranking methods also share with T-Rank~\cite{BerberichVW05} the idea of measuring freshness and activity aspects, for both individual users and their relations. 
T-Rank runs on a graph model in which nodes and edges are annotated with discrete temporal information, based on creation, deletion and modification timestamps of the items. 
A temporal window of interest and a tolerance interval are defined: the first represents the temporal range of interest to the user  (e.g., duration of an event), while the latter represents a temporal window which surrounds the window of interest (e.g., the discussion that precedes and follows the event). 
The temporal aspects of network evolution are considered through the definition of freshness and activity functions. The freshness function depends on  the time when a web page or link was last updated: it is maximal if 
a page or a hyperlink is updated with regard to the user's temporal interest, and decreases linearly with the distance to the temporal window of interest. 
The activity function reflects the rate of updates of a page's content and its incoming links,
and it is simply defined as the sum of the freshness values of modification timestamps within the tolerance interval. 
The authors define two PageRank-based algorithms, namely T-Rank and T-Rank light. The latter takes into account freshness and activity functions only for skewing the random jump probabilities during the random walk process, while T-Rank skews both the random jump probabilities and the transitions probabilities based on the temporal functions.

 However, we provide different technical solutions that  better fit   our task of lurker ranking; in particular, 
 we provide a more refined notion of activity which allows for modeling the significant variations in the lurking score trends.   
 Moreover, unlike T-Rank, we also define  cumulative  formulations of those aspects in order to  enable a time-evolving graph model for the ranking of lurkers.  In Section~\ref{sec:results},  we  have  presented an experimental comparison with T-Rank.

\subsection{User activity and interaction dynamics}
In this section we mention recent  works which, while not addressing lurking problems, are somehow related to ours since they cope with some of the topics   we have covered in Section~\ref{sec:analysis}, namely: user activity and interaction in temporal networks, time series and cluster analysis in social networks, topical evolution, newcomers, preferential attachment in directed networks, and responsiveness. 
 
User activity   can be influenced by many factors, including personal traits, communicative and social variables,  attitudes, and social influence~\cite{KrishnanA14}.   
Macropol et al.~\cite{MacropolBSPY13} considered timing and activity information available for users to uncover correlations between topic-based user activity levels and changes in sentiment. 
Since user activity trends and behavioral patterns also depend on  the structure and features of an OSN, they have often been studied in the context of specific OSNs. 
For instance, Arnaboldi et al.~\cite{ArnaboldiCPD13} examined the dynamic processes of ego networks and personal social relationships in Twitter. 
Wang et al.~\cite{WangGMZZ13} presented a detailed analysis of the dynamics of Quora, which integrates a question-answering system into an OSN, studying its user-topic graph, social graph and related questions graph. 
The relation between user activities and network structure can also determine the success or decline of an OSN, as studied  in~\cite{GarciaMS13}, where Garcia et al. analyzed the social resilience phenomenon in five online communities (i.e., Friendster, Livejournal, Facebook, Orkut and Myspace). 
Modeling engagement dynamics in social graphs is also the focus of the study by Malliaros and Vazirgiannis~\cite{MalliarosV13}. 
%
%
 
Time series analysis represents a key tool for modeling and mining temporal graphs, and has often  been used   to support   clustering or classification tasks in OSNs.  
For instance, Yang et al.~\cite{YangLY13} proposed a method for classifying Twitter users  based on the content  of their tweets. The method maps users to time series; more in detail,   tweet features are modeled as time series in order to amplify latent periodicity patterns in user communications.  
%
%
Caravelli et al.~\cite{CaravelliWSSM13} introduced a holistic dynamic clustering framework for identifying
evolving groups and   alliances across multiple time granularities in dynamic graphs.   

Considering topic modeling of social media content in addition to the time dimension, 
Wagner et al.~\cite{WagnerLPNS12} explored the impact of coupling content   with  user profile data  on the development  of the users' topical expertise.   
Hu et al.~\cite{HuSE14} 
defined a feature-based topic model and a social-based topic model 
in the context of large-scale user-generated documents available from OSN websites.  
%
%
In the context of online analysis of text streams, Saha and Sindhwani~\cite{SahaS12}  
proposed to  process incoming data together with recently seen documents over a short time window, with the purpose of  producing evolving and emerging topic sets. 
Narang et al.~\cite{NarangNMSD13} integrated  text clustering, topic similarity detection  and WordNet  
to discover time evolving conversations around topics in OSNs that do not have  explicit discussion threads (e.g., Twitter).  
 
Other specific aspects that we have analyzed in Section~\ref{sec:analysis}   concern the role of newcomers,   the responsiveness dynamics, and the preferential attachment in directed OSNs.     
Concerning  newcomers, recent attention has been paid to  the impact of community diversity in the engagement of newcomers~\cite{PanLG14}, the design of personalized engagement strategies~\cite{LopezFB12}, and  the antecedents of newcomers' participation behavior~\cite{TsaiP14}. Also, Allaho and Lee~\cite{AllahoL13} found a tendency of collaboration between masters and newcomers in software development online communities.        
Responsiveness   
is  central to gain useful insights into how users in an OSN interact with one another. 
Allaho and Lee~\cite{AllahoL14} considered the increase in responsiveness for recommending experts in collaboration networks like Github.com. 
On et al.~\cite{OnLJT13} investigated responsiveness coupled with engagingness behavior models in email networks, mainly for tasks of reply order prediction. 
Gao et al.~\cite{GaoMC15} proposed an extended reinforced Poisson process model with time mapping process to capture the dynamics underlying retweets and eventually predict the future popularity of microblogs.   
%
The preferential attachment phenomenon has long received great attention in network science. Focusing on directed networks, we observe that some studies have investigated this concept   as a growth model   for both social media  networks (e.g.,  Flickr~\cite{mislove08wosn}) and collaboration networks (e.g., Wikipedia~\cite{cap06}). 
Moreover, Kunegis et al.~\cite{KunegisBM13} performed an empirical study of preferential attachment in   OSNs for which temporal information is available. They showed that most networks follow a nonlinear preferential attachment model, whose exponent  depends on the type of network considered.

 \subsection{The challenge of latent user activities} 
 \label{sec:challenges}
In Section~\ref{sec:ddrank} we have raised an important issue in the evaluation of lurker ranking problems, which is related to the difficulty of     gathering information about latent or silent interactions among users. These are typically performed    via browsing, reading, or watching activities in the OSN environment. 

There has been a  number of relatively   recent studies that have examined latent interactions.  
By using survey data obtained by nearly 1200 recruited users, 
Burke et al.~\cite{BurkeML10} have studied user interactions on Facebook to understand relations   between  user-targeted visible actions (like those related to wall posts, comments, photo tagging, etc.), also called directed communication, and consumption actions, with loneliness and social capital bonding. They found that  directed communication is associated with higher social capital bonding and lower loneliness, whereas  social capital bridging and loneliness will increase with consumption.
In a later work~\cite{BernsteinBBK13}, the Facebook team further focused on the limitations of visible action indicators (like feedbacks and friend counts) in supporting the analysis of the size and profile of a user's audience.   

A useful tool for modeling both visible and latent activities of users is represented by \textit{clickstream} data.   Chatterjee et al.~\cite{ChatterjeeHN03} proposed to trace user clickstream data to model all activities of users, suggesting main implications in the design of OSN websites and advertisement placement.  
Benevenuto et al.~\cite{BenevenutoRCA12} provided an in-depth analysis of traffic and session patterns of user workloads. Their study is based on clickstream data that were  collected through a Brazilian OSN-aggregator website.  They examined how frequently users connect to OSN sites, how long users sessions are, how inter-request time and inter-session time data are distributed, and how the physical distance     impacts on   user interactions.   
Notably, Benevenuto et al. also discussed the opportunity of exploring clickstream data to understand silent   interactions: in this regard, they found that browsing is the most dominant behavior in Orkut (above 90\%), and that  the number of friends interacting with a user increases by an order of magnitude compared to only considering visible activities of users.  

Based on data collected from the largest OSN in China, RenRen, Jiang et al.~\cite{JiangWWSHDZ13} conducted an extensive analysis on latent interactions focusing on profile visits. 
Latent interaction graphs were found to have properties that are different from both those of social graphs and visible  interaction graphs. Latent interactions  have extremely low reciprocity, despite the fact that  RenRen   allows its users to see who recently visited their profile. 
Compared to visible interactions, latent interactions are more prevalent and more evenly distributed across a user's friends.  A significant part of profile visits comes from non-friends of a user, while  the majority of visitors do not browse the same profile twice.  Moreover, profile popularity does not seem to be strictly correlated with the frequency of updates.    

We envisage that the lessons learned from the above mentioned studies concerning latent activities of OSN users, could be helpful to enhance the understanding of  lurking  behaviors. In particular, involving information on latent activities extracted from profile visits and clickstream data, would pave the way  to new opportunities of evaluation of lurker ranking  problems. In this regard, we would like to stress again that our time-aware LurkerRank methods can equally deal with visible as well as latent information-consumption activities, and that they are suitable for new scenarios of data-driven ranking evaluation.

\section{Conclusion} 
 \label{sec:conclusion}
  
  In this work, we   advanced research on lurking in OSNs in a twofold manner.    We studied   the dynamics of lurking behaviors in OSNs, by performing   a rigorous analysis aimed to understand how lurkers relate to other types of users and how patterns of lurking behaviors evolve over time. More in detail, we   compared lurkers and  inactive users as well as   lurkers and newcomers,  investigated  preferential attachment between lurkers and active users,    studied   the lurkers' responsiveness to others' actions, performed a cluster analysis of the lurking trends over time  and a  topic-sensitive  analysis of evolving patterns of lurking behavior.  
We also overcome the time-related limitation of previous formulations of lurker ranking methods. In this regard, we  developed    measures related to   freshness and     activity trend, both for individual users and for interactions between users. Such measures were used as key elements in   time-aware LurkerRank methods,  following either a time-static or a time-evolving graph model.  Results have shown the significance of our time-aware LurkerRank methods.  We have finally discussed open issues, concerning   new opportunities of evaluation of  lurker ranking problems based on the exploitation of user latent activities.

\end{document}